\documentclass[modern]{aastex61}

\usepackage{longtable}
\usepackage{amssymb} 

\accepted{January 15, 2019}

\submitjournal{ApJ}

\shorttitle{Brown dwarf rotation-disk connection}
\shortauthors{Moore et al.}

\begin{document}

\title{The Rotation-Disk Connection in Young Brown Dwarfs: Strong Evidence for Early Rotational Braking}

\correspondingauthor{Keavin Moore}
\email{keavin.moore@gmail.com}

\author{Keavin Moore}
\affiliation{Faculty of Science, York University, 4700 Keele Street, Toronto, ON M3J 1P3, Canada}
\affiliation{Department of Earth \& Planetary Sciences, McGill University, 3450 rue University, Montr{\'e}al, QC H3A 0E8, Canada}

\author{Aleks Scholz}
\affil{SUPA, School of Physics \& Astronomy, University of St Andrews, North Haugh, St Andrews, KY16 9SS, United Kingdom}

\author{Ray Jayawardhana}
\affiliation{Faculty of Science, York University, 4700 Keele Street, Toronto, ON M3J 1P3, Canada}
\affiliation{Department of Astronomy, Cornell University, Ithaca, NY 14853, USA}

\begin{abstract}
We use Kepler/K2 lightcurves to measure rotation periods of brown dwarfs and very low mass stars in the Upper Scorpius star-forming region. Our sample comprises a total of 104 periods. Depending on the assumed age of Upper Scorpius, about a third of them are for brown dwarfs. The median period is 1.28\,d for the full sample and 0.84\,d for the probable brown dwarfs. With this period sample, we find compelling evidence for early rotational braking in brown dwarfs, caused by the interaction between the central object and the disk. The median period for objects with disks is at least 50\% longer than for those without. Two brown dwarfs show direct signs of `disk-locking' in their lightcurves, in the form of dips that recur on a timescale similar to the rotation period. Comparing the period samples for brown dwarfs at different ages, there is a clear need to include rotational braking into period evolution tracks between 1 and 10\,Myr. A locked period over several Myr followed by spin-up due to contraction fits the observational data. We conclude that young brown dwarfs are affected by the same rotational regulation as stars, though they start off with significantly faster rotation, presumably set by initial conditions.
\end{abstract}


\section{Introduction} \label{sec:intro}

Low-mass stars lose angular momentum over the course of their life, through rotational braking via the star-disk interaction in the first few Myr and via magnetically driven stellar winds on longer timescales \citep{herbst07}. How effective these processes are in objects with ultra-low masses, i.e. brown dwarfs and free-floating planets, remains an open question that is the subject of active research \citep{zhou16,schwarz16,vasconcelos17,bryan18}. There is a consensus, however, that brown dwarfs rotate faster than solar-mass stars at all ages \citep[see review by][and references therein]{bouvier14}. For very young brown dwarfs, rotation periods are in the range of 1-3\,days \citep{scholz04,mohanty05}, whereas evolved brown dwarfs spin with periods of a few hours up to a day \citep{reiners08,metchev15,zapatero06}. 

The ideal way to assess rotation rates for large samples is measuring periods from photometric lightcurves. Through rotation, surface features like magnetic spots or dusty clouds will modulate the brightness, with a period corresponding to the rotation period. Measuring rotation periods for objects with a range of ages is the fundament for analysing the evolution of angular momentum in the substellar regime. First periods for brown dwarfs ($M<0.08\,M_{\odot}$) in star-forming regions and open clusters have been derived from ground-based wide-field data over the past 15 or so years, including $\sigma$\,Orionis \citep{scholz04,caballero04,cody10}, Chamaeleon-I \citep{joergens03,codyhill14}, NGC2264 \citep{lamm05}, and ONC \citep{rodriguez09}. In addition, targeted observations of individual field brown dwarfs have yielded substantial period samples \citep[e.g.][]{bailer01,harding13,metchev15,miles17}. For brown dwarfs with typical periods around a day and small photometric amplitudes, ground-based monitoring --with daytime gaps and red noise caused by atmospheric effects-- remains a challenging proposition.

The failing of the reaction wheels on the Kepler satellite and its ensuing `second' mission K2 was a blessing in disguise for work on stellar and substellar rotational evolution. During its K2 incarnation, the spacecraft has been able to observe numerous young clusters near the ecliptic with unprecedented cadence and precision \citep{howell14}, allowing us to revisit the issue of brown dwarf rotation with significantly improved data. This paper is the third in a series aiming to establish robust period samples for well-characterised samples of brown dwarfs, using K2 lightcurves. 

In the first paper, we published the first periods for a (small) sample of brown dwarfs in the $\sim 10$\,Myr old Upper Scorpius region \citep{scholz15}, suggesting that disk locking is inefficient in the substellar regime. In the second paper, we used multiple period search techniques to establish periods for 18 spectroscopically confirmed Taurus brown dwarfs \citep{scholz18}. We found a link between the presence of disks (detected via Spitzer infrared excess) and slow rotation, thus evidence for rotational braking by the disks. Moreover, when extrapolated to the age of the solar system, the periods of young brown dwarfs fit the spin-mass trend of solar system planets, demonstrating the link between planetary, substellar, and stellar rotation. 

In this paper we return to Upper Scorpius, now making use of the full sample of brown dwarfs covered in K2 campaigns 2 and 15, as well as the improved systematics correction compared to 2015. The main purpose of this paper is to examine the rotational regulation in brown dwarfs by disks with the enlarged sample and improved methodology. Broadly speaking, there are three different ways of testing the link between rotation and disks: a) comparing period distributions for objects with and without disks, identified by excess mid-infrared emission \citep{rebull06}, b) finding objects with direct evidence for a disk feature (e.g., a warp) that is co-rotating with the surface \citep{stauffer15}, and c) comparing rotation rates at different ages to test for rotational braking. In analysing our new period sample in UpperSco, we find that all three approaches consistently yield evidence for rotational regulation by disks.

\section{The Sample} 
\label{sec:samp}

The Upper Scorpius star-forming region has been observed twice during the course of the \textit{Kepler}/\textit{K2} mission \citep{howell14}. Campaign 2 lasted 82 days, from August 23, 2014 to November 13, 2014, while Campaign 15 occurred over 89 days, from August 23, 2017 to November 20, 2017. 

Our sample of brown dwarfs and very low mass stars was constructed as follows. We start with the 52 objects examined in \citet{scholz15}. This sample has been selected using multi-band photometry and proper motions by \citet{dawson13}. Spectroscopic follow-up has shown that the overwhelming majority of these sources are mid to late M dwarfs with evidence of youth \citep{dawson14}. We select 9 more objects observed in Campaign 2, and 8 from Campaign 15, found by cross-matching the membership list by \citet{cook17} (more precisely, their L-ZYJHK and C-ZYJHK HK-cut samples) and the full list of K2 targets in those campaigns. Thus, our primary sample in the current paper comprises a total of 69 objects, the majority of which have been  confirmed spectroscopically (see Table \ref{tab:samplea_periods}). We refer to this primary sample as `sample A'.

\begin{figure}[htb]
\includegraphics[width=16cm]{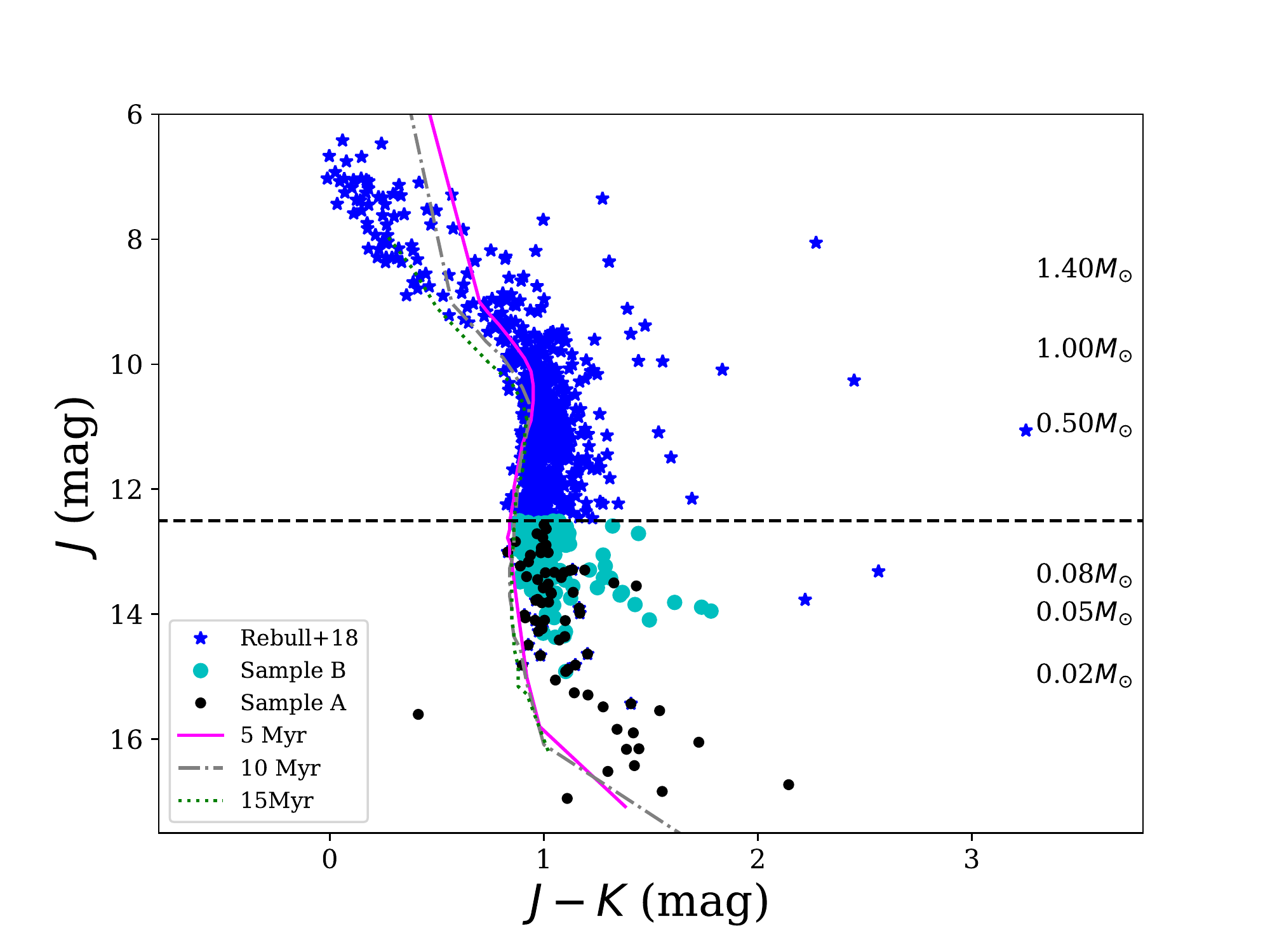}
\caption{\label{fig:sampsel} $(J, J-K)$ colour-magnitude diagram, displaying our sample selection. The 5 Myr, 10 Myr, and 15 Myr isochrones are taken from \citet{baraffe15}, shifted from absolute magnitude to apparent magnitude using a distance to UpperSco of 145\,pc \citep{cook17}. Objects from \citet{rebull18} are shown as blue star symbols, while our A and B samples are displayed in black and cyan, respectively.}
\end{figure}

Sample A only includes objects from a fraction of the entire UpperSco association. \citet{rebull18} have recently compiled a list of likely members based on photometry and proper motions for a much larger region. We supplement our object list by selecting a subsample from \citet{rebull18} using a magnitude cutoff $J>12.5$ and a colour cutoff $J-K<2$. These cutoffs are defined based on our sample A and select objects with little extinction and masses near or below the substellar boundary (see Figure \ref{fig:sampsel}). We also largely avoid the area of the sky including the much younger star forming region $\rho$\,Oph ($246.14<$RA$<247.43$, $-25.19<$Dec$<-24$) to prevent contamination from this region. This step gives us 135 more objects, which we refer to as `sample B'. In total, our A and B samples contain 204 objects. Because of their different selection criteria, we keep these two samples separate for the analysis in this section, but we find that in terms of their rotational properties there is no significant difference. 

Fig.~\ref{fig:sampsel} is a colour-magnitude diagram, displaying $J$ vs. $J-K$ for samples A (black) and B (cyan). This figure includes all objects from \citet{rebull18} as blue star symbols (excluding the $\rho$\,Oph region), with 5, 10, and 15 Myr isochrones from \citet{baraffe15} overplotted. The adopted cut-off of $J > 12.5$ is shown as the dashed black line, and our A and B samples are displayed below this cut-off. Approximate mass limits are indicated, based on $J$ magnitudes from the 10\,Myr isochrone of \citet{baraffe15}. Assuming an average distance of 145\,pc \citep{cook17}, the substellar limit ($0.08\,M_{\odot}$) in UpperSco should be between $J=12.8$ for an age of 5\,Myr and 13.5 for an age of 10\,Myr. Our selection should safely encompass all substellar objects, but will also include a number of very low mass stars with masses above the substellar limit.

\section{K2 Lightcurves and Period Search} 
\label{sec:lc}

\subsection{K2 Lightcurves}

The main purpose of this study is the search for lightcurves with periodic sinusoidal variations, which can be interpreted as rotation periods caused by surface spots on the targets. We use the high-level science product K2SFF reduced lightcurves (based on an algorithm from \citet{vanderburg14}) from the K2 MAST archive for investigation of periodicity, for both campaign 2 and campaign 15. K2 lightcurves suffer from systematics due to spacecraft drift, and a number of algorithms have been developed to correct for that, K2SFF being one. The suitability of K2SFF lightcurves to recover periodic signals in comparison with other algorithms has been demonstrated by, for example, \citet{esselstein18} and also in our own previous study \citep{scholz18}. In agreement with the literature (see for example \citet{rebull18}) we find by visual inspection that lightcurves from different algorithms generally give consistent results when searching for coherent periods that are stable over many cycles (as expected for our targets).

The lightcurves for the 69 (sample A) plus 135 (sample B) targets were first examined by eye. Many have obvious periods throughout the observations. There is a clear `jump' present in many lightcurves around the middle of the campaign; this step function between the first and second half of the campaign has been noted before as an artifact of the K2SFF reduction process (e.g. \citealt{hedges18}), as the lightcurve is detrended in separate `stability regimes' to avoid the reorientation of the spacecraft. Many of our lightcurves also show a large spike in flux near the middle of the campaign. The pixel data identify these events as either `argabrightening', in which all pixels are equally illuminated, or `cosmic ray in collateral data'. It appears that Mars crosses the frame during this time in the campaign, adding flux across all pixels of the CCD, resulting in an artifact after K2SFF reduction. Therefore, any anomalies occurring around this time are discarded. 


\subsection{Period Search}

We calculated the autocorrelation function (ACF) of all 204 lightcurves to measure periods. The autocorrelation function, as the name suggests, tests the correlation of a lightcurve with itself, shifting the copy by a time delay $\delta$. The ACF will then show a peak at $\delta = 0$, as expected, and any additional peaks at $\delta = N \times P$ reveal an underlying period, if one is present, allowing its extraction. The ACF is insensitive to the shape of the periodic signal and robust against changes in amplitude. It has been widely adopted as one of the main period search methods in studies using Kepler and K2 data \citep{mcquillan13,aigrain15,stelzer16,giles17}.

Our procedure was analogous to that of \citet{scholz18}, computing the ACF first for the entire lightcurve, and then independently for 7 equal segments of the lightcurve. A period was accepted if present in at least 2 of 7 segments within a tolerance of 0.1\,d, and if it could be visually confirmed by eye. The standard deviation over the segments with a consistent period gives us an estimate for the error of the period. The few ACF periods which could not be convincingly visually confirmed in their corresponding segments, even if appearing in multiple segments, had only weak maxima in their ACF. As a final step, we phase-fold a segment of the lightcurve. These phase plots are shown in the Appendix. For clarity, we include Fig.~\ref{fig:search_process} showing the steps of our period search as described above.


\begin{figure}[t]
\centering
\includegraphics[width=1.0\textwidth]{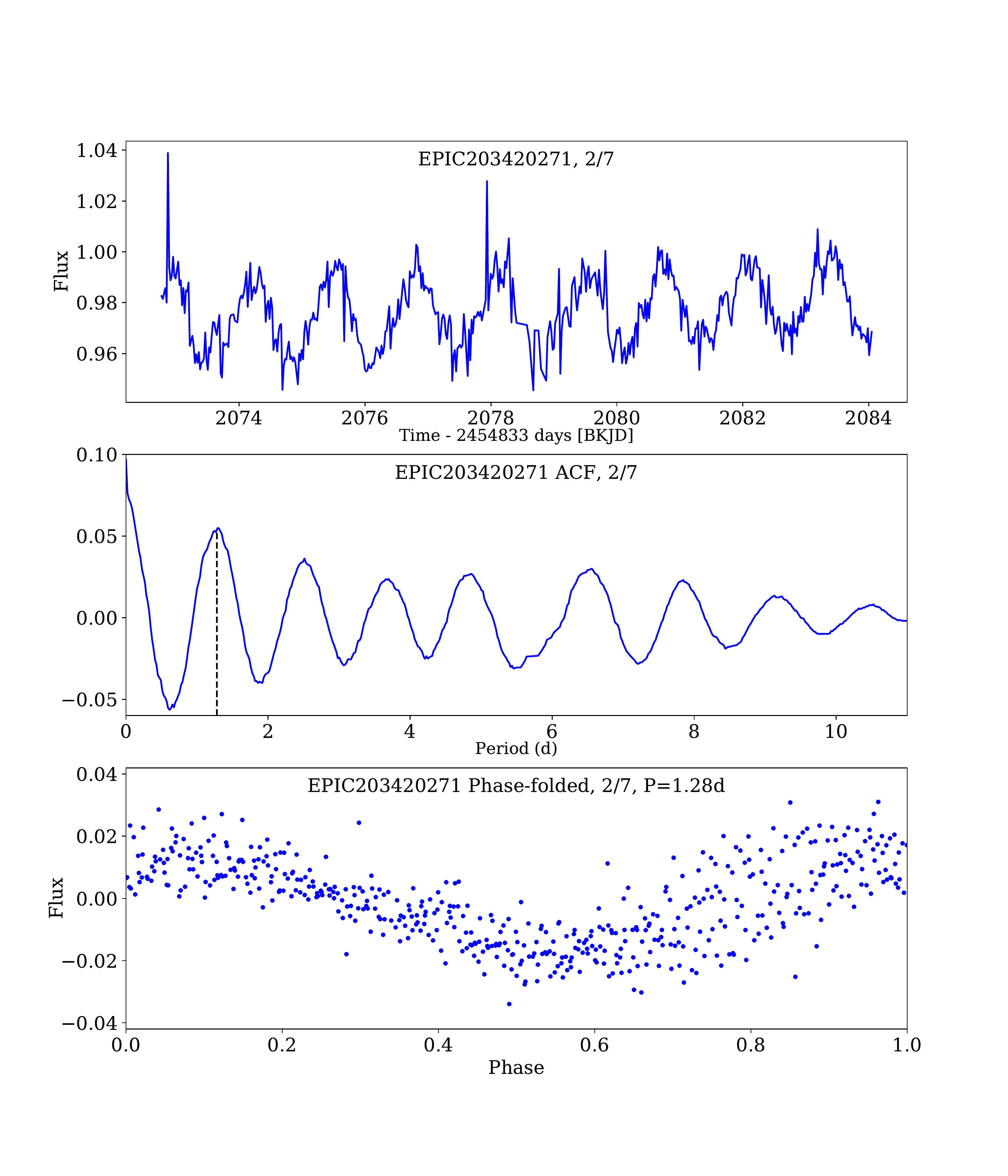}
\caption{\label{fig:search_process} Our period search process, illustrated here for EPIC203420271. After visual inspection of the entire lightcurve, we note that there is a periodic signal. We then divide the lightcurve into 7 segments, all of which show periodicity. The top panel shows the second of these seven segments. The calculated autocorrelation function is shown in the middle panel, with a clear peak at zero, and a strong peak at 1.28\,d. The bottom panel shows the phase-folded lightcurve, after subtraction of the average flux, using the identified period of 1.28\,d from the ACF.}
\end{figure}

In sample A, we measure a total of 25 periods. 22 (88\%) of them were consistent in 5 or more segments, while the remaining 3 (12\%) were found in 3 or 4 segments of the lightcurve. In sample B, 79 periods were detected. 65 of them (82.3\%) were consistent in 5, 6, or 7 segments, while the other 14 (17.7\%) were found in 2, 3, or 4 segments. Our 25 accepted periods for sample A are shown in Table~\ref{tab:samplea_periods}, with uncertainties calculated as the standard deviation between segments for which the period was confirmed. Sample B adds another 79 periods, which are shown in Table~\ref{tab:sampleb_periods}. Our period search, then, resulted in period measurements for 104 out of the 204 objects altogether. We show phased lightcurves for all 104 periodic objects in Figs.~\ref{fig:phases_a1} to ~\ref{fig:phases_b5}. 


\begin{center}
\begin{longtable}[t]{l|ccccccc}
\caption{Our catalogue of objects with periods in sample A, including their spectral type, where available. The period, the number of segments with the same period, and its uncertainty is listed, followed by the proposal ID for each object. \label{tab:samplea_periods}} \\
\hline
\hline
EPIC & RA & Dec & SpT & $P_{ACF}$  & $N$   & $\Delta P$ & Proposal ID \\
     & (J2000) & (J2000) & & (d)     &    & (d)     &  \\
\hline
\textit{K2 GO2010} & & & & & & & \\
202632400 & 16 17 56.087 & -28 56 39.97 & & 1.38 & 3 & 0.05 & GO2010 \\
203083616 & 16 14 52.534 & -27 18 55.71 & & 0.65 & 6 & 0.01 & GO2010 \\
203348744 & 16 03 02.358 & -26 26 16.37 & & 4.29 & 3 & 0.02 & GO2010 \\
203420271 & 16 03 37.991 & -26 11 54.43 & & 1.28 & 7 & 0.04 & GO2010 \\
203544427 & 15 55 42.290 & -25 46 47.79 & & 1.12 & 6 & 0.03 & GO2010 \\
204078097 & 16 09 58.525 & -23 45 18.61 & M6.5$^2$ & 1.41 & 6 & 0.09 & GO2010 \\
204099713 & 16 11 26.300 & -23 40 05.97 & M5.5$^4$ & 1.77 & 5 & 0.03 & GO2010 \\
204126288 & 16 16 45.394 & -23 33 41.39 & M5.0$^4$ & 0.21 & 7 & 0.01 & GO2010 \\
204149252 & 16 13 34.766 & -23 28 15.61 & M5.75$^3$ & 0.97 & 7 & 0.04 & GO2010 \\
204202443 & 16 15 28.195 & -23 15 43.95 & M5.5$^3$ & 1.03 & 5 & 0.03 & GO2010 \\
204204414 & 16 15 36.479 & -23 15 17.52 & M5.75$^3$ & 0.61 & 7 & 0.01 & GO2010 \\
204239143 & 16 11 38.375 & -23 07 07.27 & M6.25$^3$ & 0.71 & 7 & 0.03 & GO2010\\
204344180 & 16 14 32.870 & -22 42 13.35 & M6.5$^4$ & 1.76 & 6 & 0.04 & GO2010 \\
204367193 & 16 11 54.395 & -22 36 49.19 & M6.25$^3$ & 0.48 & 7 & 0.01 & GO2010 \\
204393705 & 16 13 26.656 & -22 30 34.84 & M6.25$^3$ & 1.48 & 7 & 0.02 & GO2010 \\
204418005 & 16 09 04.514 & -22 24 52.39 & M7.0$^4$ & 0.54 & 6 & 0.03 & GO2010 \\
204439854 & 16 11 34.703 & -22 19 44.21 & M5.75$^3$ & 0.60 & 4 & 0.02 & GO2010 \\
204451272 & 16 08 22.294 & -22 17 02.90 & M5.75$^3$ & 0.96 & 6 & 0.01 & GO2010 \\
204555809 & 16 09 01.976 & -21 51 22.54 & & 0.27 & 7 & 0.01 & GO2010 \\
\hline
\textit{K2 C02} & & & & & & \\
202795175 & 16 23 41.870 & -28 20 12.70 & & 0.83 & 6 & 0.04 & GO2063 \\
203004488 & 16 18 44.316 & -27 35 30.40 & & 1.35 & 5 & 0.08 & GO2063 \\
204250417 & 16 15 13.612 & -23 04 26.12 & M6.5$^4$ & 1.28 & 7 & 0.10 & GO2045 \\
\hline
\textit{K2 C15} & & & & & &  \\
203590915 & 15 50 59.935 & -25 37 11.66 & & 0.76 & 6 & 0.03 & GO15043 \\
249146655 & 15 42 08.305 & -26 21 13.84 & M5.7$^1$ & 0.38 & 7 & 0.02 & GO15043 \\
249202312 & 15 41 55.626 & -25 38 46.53 & & 1.42 & 7 & 0.03 & GO15043 \\
\hline
\multicolumn{8}{l}{RA, Dec, and proposal ID from K2 MAST Archive.} \\
\multicolumn{8}{l}{$^1$: \citet{dawson14}, $^2$: \citet{bm09}, $^3$: \citet{ldh11},} \\
\multicolumn{8}{l}{$^4$: \citet{shc08}.} \\
\end{longtable}
\end{center}

\subsection{Comparison with Literature}

\begin{figure}[t]
\centering
\includegraphics[width=1.0\textwidth]{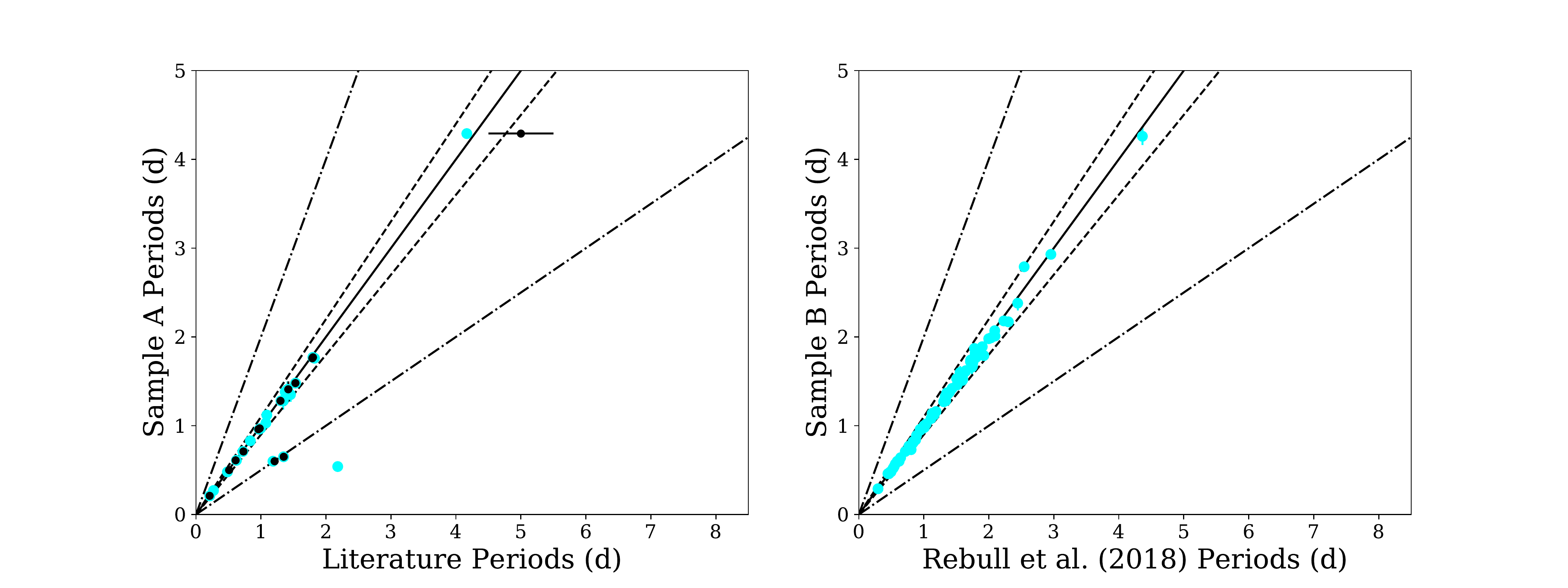}
\caption{\label{fig:per_corr} A comparison of the periods for sample A (left panel) and B (right panel) with those in the literature. In the left panel, the black circles indicate comparisons with \citet{scholz15, cody18, hedges18}, while, in both panels, cyan circles indicate comparisons with \citet{rebull18}. It is clear that there is a strong correlation between our measured periods and those in the literature. The black dashed line in both panels indicates $y = x$, i.e., a perfect agreement, while the black dashed lines show $y = 1.1 x$ and $y = 0.9 x$, a 10\% discrepancy, and the black dot-dashed lines show $y = 2x$ and $y = 0.5x$, i.e. period disagreeing by a factor of two. Uncertainties included for \citet{scholz15}, but not published in \citet{cody18}, \citet{hedges18}, or \citet{rebull18}.}
\end{figure}

In Fig.~\ref{fig:per_corr} we show our periods in comparison with previously published measurements in the literature, for A and B samples separately. 

For sample A, we independently recover 13 periods from \citet{scholz15}, including their noted outlier EPIC203348744, which they measured as $P \sim 5 $ d; for the other 3 objects, a period could not be adequately determined from the K2SFF lightcurve using ACF and visual inspection, and thus they are excluded from our sample. Our periods are in agreement, within the uncertainties, with \citet{scholz15} with the exception of EPIC204439854, where we find a period half their measured value. A period was previously published for EPIC203083616 \citep{cody18, hedges18}; our value is half that of their measurement. These 14 objects are shown as black circles in the left panel of Fig.~\ref{fig:per_corr}. In this figure, we show the 1:1 relation (solid line), the lines corresponding to 10\% deviation (dashed) and for factor-of-two discrepancy (dot-dashed).

Twenty-two of the objects from sample A and all 79 from sample B have measured periods in \citet{rebull18}, shown in Fig.~\ref{fig:per_corr} as well as cyan circles. Most (98/101) Rebull periods are in agreement (within 10\%) with our measured periods, between the two samples. For the remaining three objects, the periods are roughly multiples of ours: for EPIC203083616 and EPIC204439854, our measured period is half that measured by Rebull, and for EPIC204418005, it is one-quarter the literature value. In contrast to the ACF-based period search on segmented lightcurves used here, \citet{rebull18} apply Lomb-Scargle periodograms on the entire lightcurve. In summary, the comparison with the literature confirms the validity of our period sample.

\subsection{Injected periods}

\begin{figure}[t]
\centering
\includegraphics[width=0.8\textwidth]{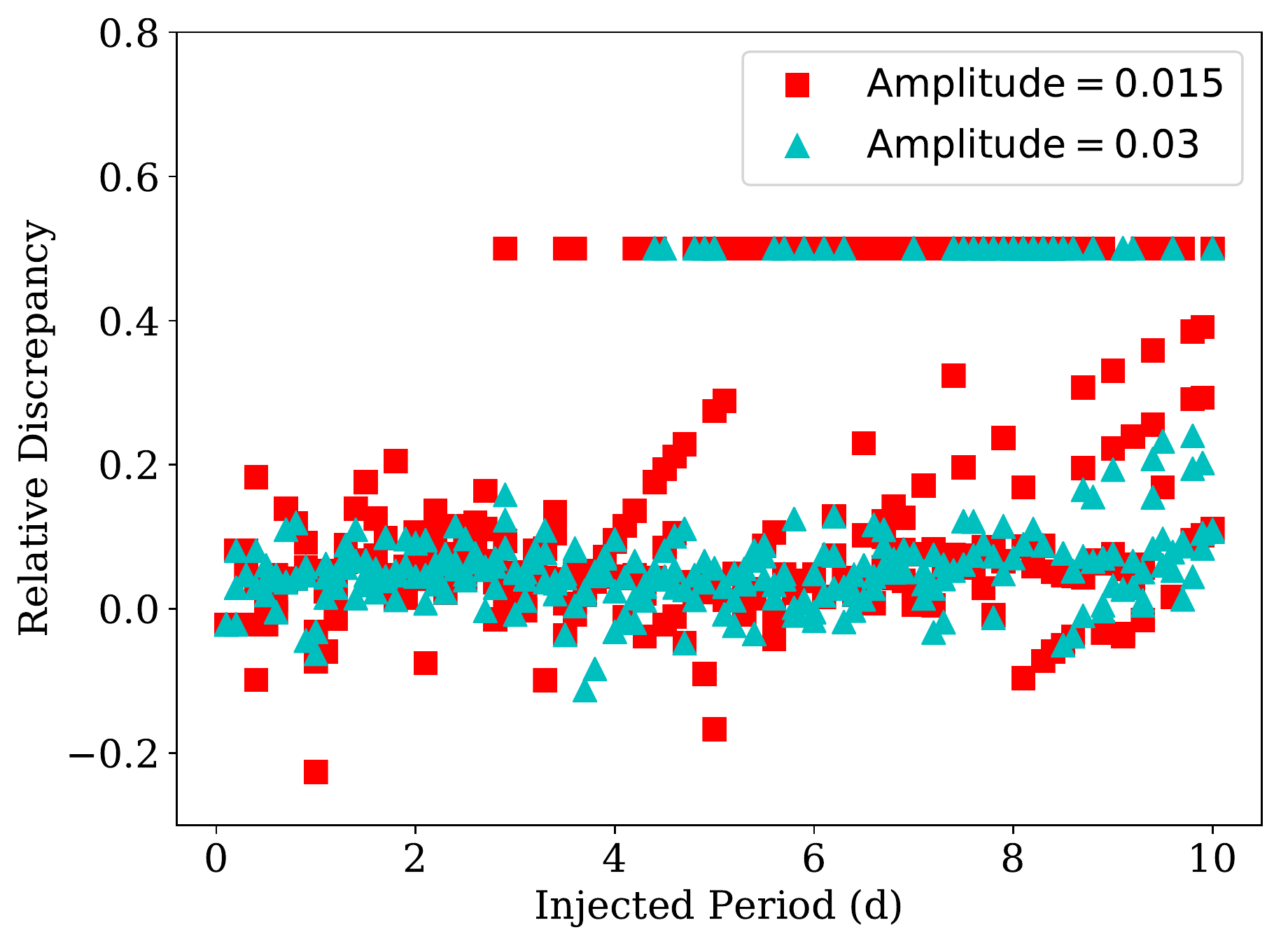}
\caption{\label{fig:injected} A plot of relative discrepancy vs. injected period, for our one hundred injected periods and two different amplitudes in three different lightcurves without a measured period. Here, we define relative discrepancy as (injected period - recovered period)/injected period. It is clear that the majority of our ACF-measured values are within 10\% of the injected value. See text for more details.}
\end{figure}

We do not find any periods $\lesssim 0.20$\,d or $\gtrsim 5$\,d in our ACF period search of K2SFF lightcurves. To test the validity of these limits, we create datasets with one hundred different sinusoidal periods (0.1 d to 10 d, in steps of 0.1 d), each with two different amplitudes (0.015 and 0.03, measured from zero to peak), typical for the periodicities that we found. We inject these periods into non-periodic lightcurves from three different objects, to retain the noise and sampling of K2. These three objects cover the range of $J$-band magnitudes seen in our sample: EPIC202978875 ($J = 12.939$), EPIC202633073 ($J = 14.097$), and EPIC204341806 ($J = 16.947$); this provides a sample of 600 injected periods. We then perform a period search using our ACF process, dividing each injected lightcurve into 7 segments. A period is recovered if it is found in multiple segments with $<0.1$\,d tolerance. Due to the large number of injected periods, we do not perform the visual examination as we did during the period search. Fig.~\ref{fig:injected} shows the relative discrepancy between injected and recovered period vs. the injected period. Objects for which a period was not recovered are plotted at a relative discrepancy of 0.5; for the smaller amplitude, this corresponds to 68 points, and 40 points for the larger amplitude. In total, 108/600 (18\%) of the injected periods are not recovered. We successfully recover 173 periods for the smaller amplitude, and 198 for the larger amplitude. Overall, 371/600 or 61.8\% of injected periods are recovered. For the remaining 121 test lightcurves we find a period, but not within 10\% of the injected period. In summary, the injection test demonstrates that the method is sensitive to a wide range of periods, although we will be incomplete for amplitudes of $<0.015$ and for periods $>4$\,d.

\subsection{Period distributions}

\begin{figure}[t]
\centering
\includegraphics[width=1.0\textwidth]{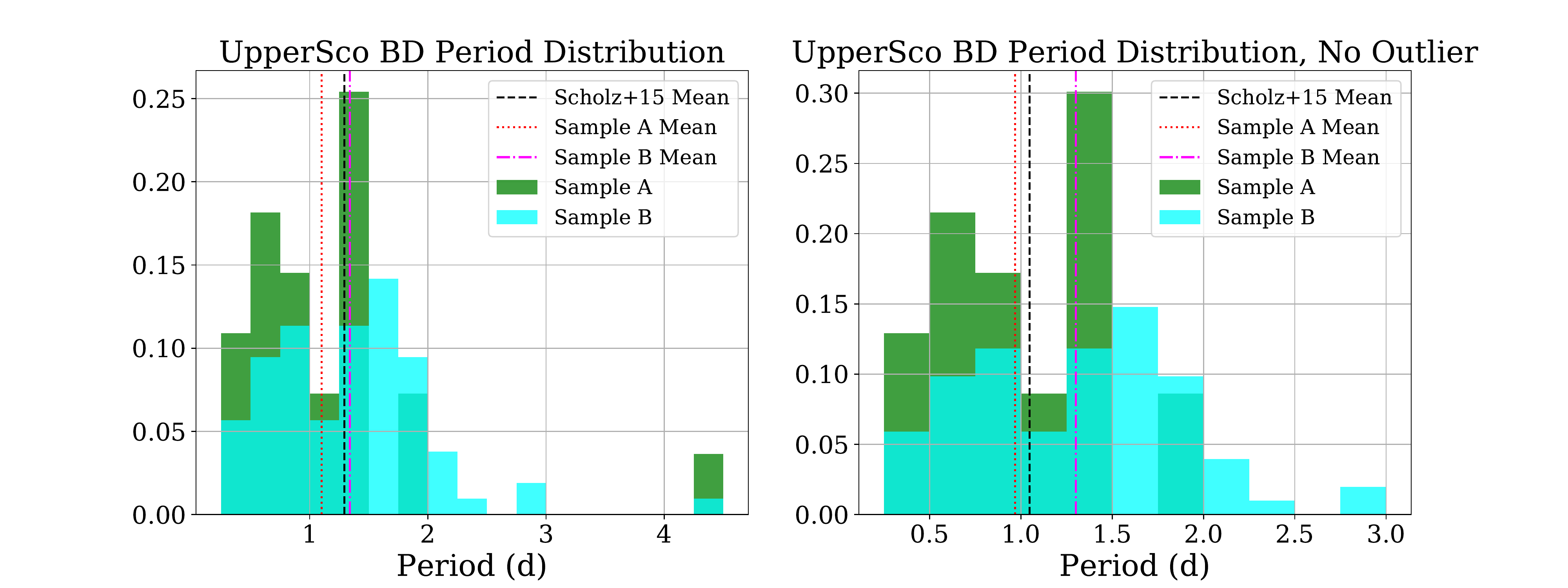}
\caption{\label{fig:hist20_all} Period distribution of sample A (green) and B (cyan). The left panel shows all periods from our catalogue, and the right panel shows the distribution after removal of the single long-period outlier. The red dashed lines are the average of sample A  (with and without the outlier), the magenta lines are the average of sample B periods (also with and without the outlier). For comparison, the black dashed lines are the average of the 16 periods (or 15, after removal of outlier, in the right panel) from \citet{scholz15}.}
\end{figure}

Fig.~\ref{fig:hist20_all} shows the histograms of the period distributions for sample A (green) and sample B (cyan), with and without the single long-period outlier at 4.29 and 4.26\,d, respectively. The dashed red and magenta lines denote the average of sample A and B respectively, while the black dashed line represents the average for the sample from \citet{scholz15}. The periods for sample A range from 0.21\,d to 1.77\,d (excluding the aforementioned outlier), with a mean and median of 1.10 and 0.97\,d (0.97\,d and 0.97\,d, respectively, without the outlier). For sample B, the range (again, excluding its outlier) for the period distribution is from 0.29 to 2.93\,d; the mean and median are 1.34 and 1.33\,d, respectively (1.30\,d and 1.33\,d without the outlier). Our period samples broadly confirm the previously found distribution, but expands the sample size considerably. The similar means and medians also suggests that there are no systematic differences in the period distributions for samples A and B. For the full sample of 104 periods, the median is 1.28\,d. Selecting only the probable brown dwarfs adopting $J=13.3$ as a threshold, the median drops to 0.84\,d. Thus, with very few exceptions, brown dwarfs at the age of UpperSco are fast rotators with typical periods around 1\,d and with very few exceptions below 3\,d. All this is in line with findings by other authors \citep{rodriguez09}.

\subsection{Brown dwarfs vs. stars}

\begin{figure}[htb]
\centering
\includegraphics[width=1.0\textwidth]{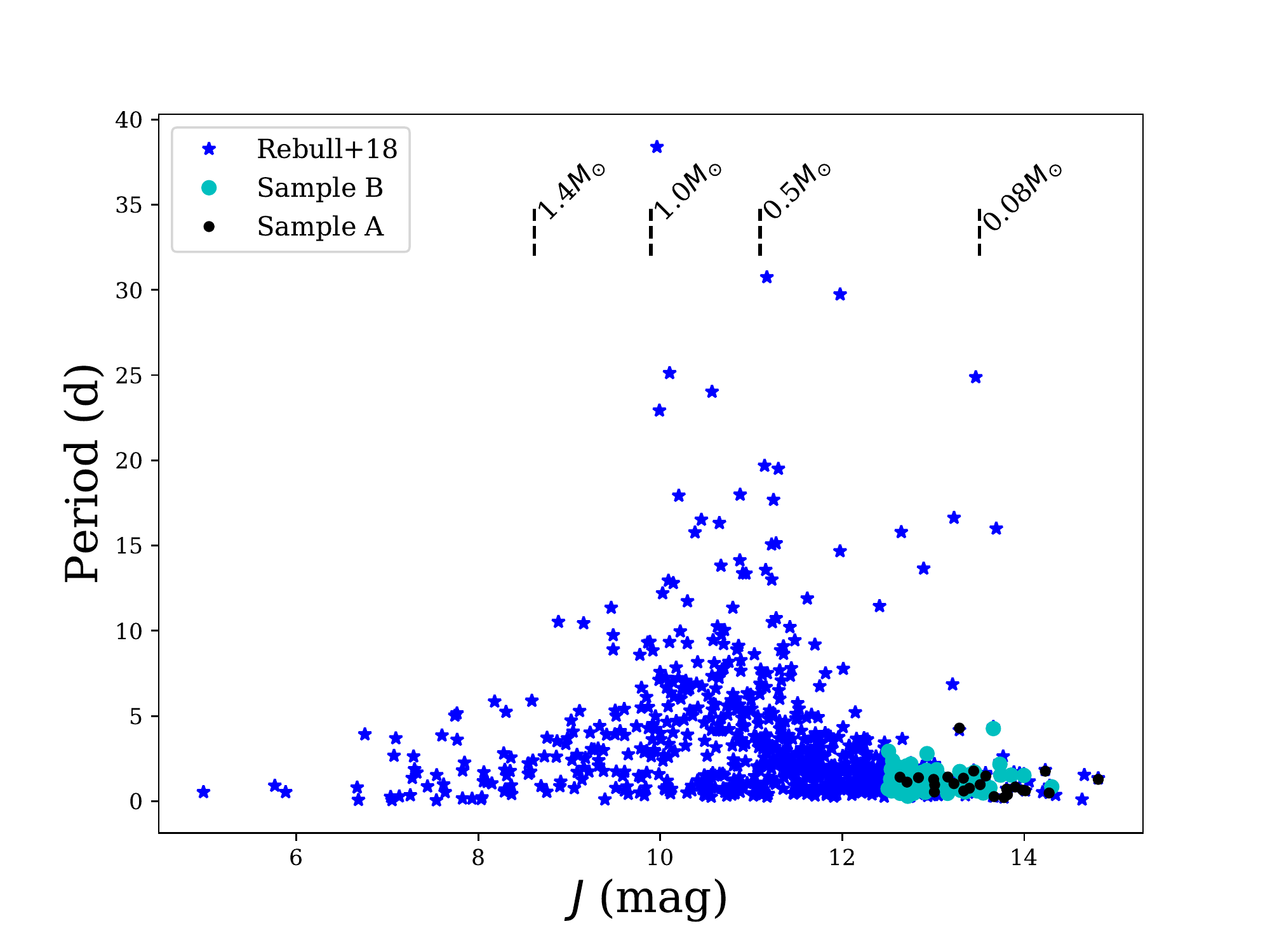}
\caption{\label{fig:per_comparison} A comparison of Upper Scorpius stellar periods \citep{rebull18} (blue star symbols) and brown dwarf periods (our sample A, black circles, and sample B, cyan circles) plotted against 2MASS $J$ magnitude. Mass limits determined by converting $M_J$ from the 10 Myr isochrones of \citet{baraffe15} to apparent magnitude $J$, using a distance of 145\,pc.}
\end{figure}

Fig.~\ref{fig:per_comparison} shows the UpperSco periods from \citet{rebull18} and our measured brown dwarf periods, from both  A and B samples, vs. 2MASS $J$ magnitude. Again we excluded the younger $\rho$\,Oph region (see Sect.~\ref{sec:samp}). As expected from our sample selection, the targets are all on the faint end of this diagram, and are expected to have masses below or around the substellar boundary. In contrast, the \citet{rebull18} sample includes the entire stellar mass range. Assuming that the objects are roughly coeval and not significantly affected by extinction, the J-band magnitude serves as a proxy for mass. Approximate mass limits are overplotted in Fig.~\ref{fig:per_comparison} from the \citet{baraffe15} tracks and adopting a distance of 145\,pc.

Similar to previously studied star-forming regions, the periods in UpperSco show a strong mass dependence. The average period and the range of periods drop significantly with mass, as found previously in the ONC \citep{rodriguez09} and NGC2264 \citep{lamm05}. The mean period in our samples is 1.10\,d, whereas the average in the periods for stars in UpperSco is more than three times as long (3.38\,d). While brown dwarf periods predominantly cluster below 3\,d, more massive stars show a much wider range of periods up to several weeks.

\begin{figure}[t]
\centering
\includegraphics[width=1.0\textwidth]{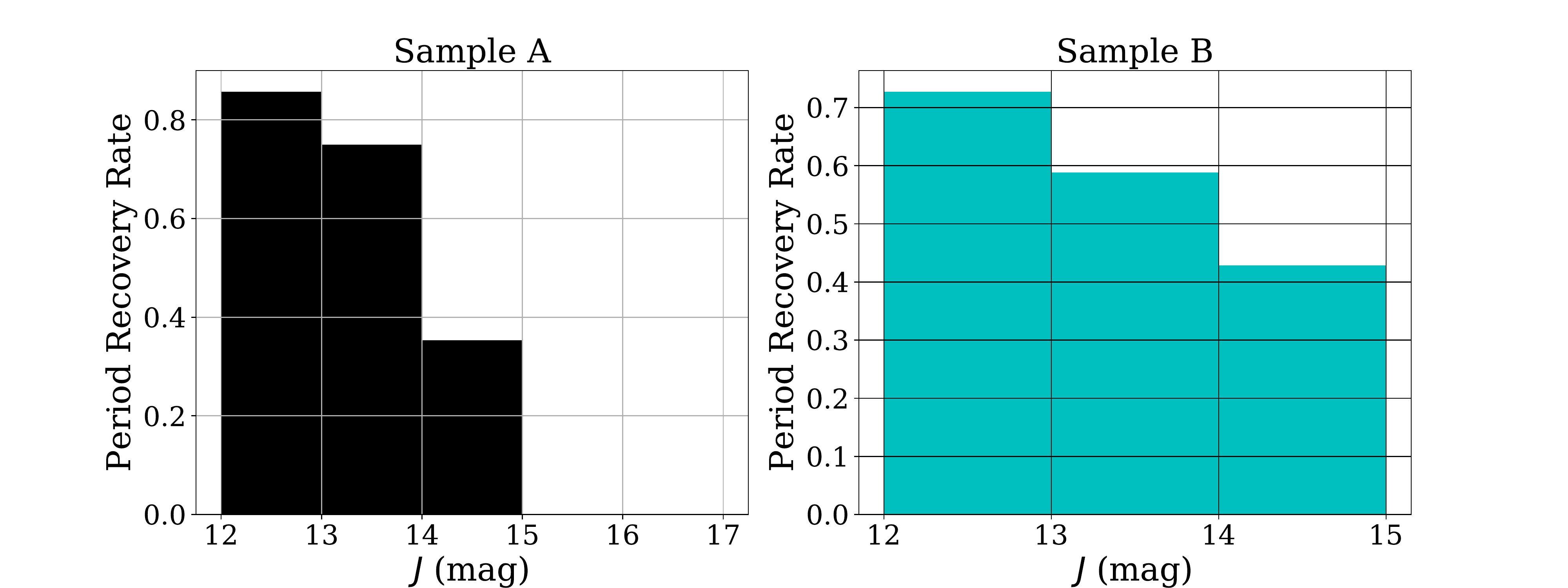}
\caption{\label{fig:recoveryrate} Period recovery rates of sample A (left panel, black) and sample B (right panel, cyan) as a function of their $J$-band magnitude. We divide the samples into bins of width $J = 1$; sample A contains objects with $J \sim 12-17$, while sample B contains objects with $J \sim 12-15$. As can be seen, we recover significantly more periods for brighter objects, with a steady decrease as the objects become fainter. Note that, although sample A contains objects in the $J=15-17$ bins, we do not recover any periods in that magnitude region.}
\end{figure}

Fig.~\ref{fig:per_comparison} also reveals that we do not find periods for the low-mass brown dwarfs in the samples: whereas the magnitude distribution of the samples extends to $J\sim 17$ and 0.02$\,M_{\odot}$, the period sample has a lower limit at $J<15$. Thus, the period sample is dominated by high-mass brown dwarfs and very low mass stars. This is further illustrated by Fig.~\ref{fig:recoveryrate} which shows the recovery rate for periods as a function of magnitude. For sample A, we only recover periods for objects up to $J=15$, mostly for the brightest objects between $J=12-13$. While sample A includes objects in the $J>15$ bins (up to $J \sim 17$), we do not find any periods for these fainter objects. Since sample B only includes objects with $J<15$, we recover periods in all magnitude bins. Future observations with greater depth should aim to measure rotation periods for the faint objects in this region.

\section{Disks vs. rotation}
\label{sec:disks}


\subsection{Identifying disks}

To test for the presence of disks, we compile photometry from the Wide-field Infrared Survey Explorer (WISE) \citep{cutri13} in the channels $W1$ to $W4$, corresponding to wavelengths of 3.4, 4.6, 12, and 22 $\mu$m. While longer wavelengths such as $W3$ and $W4$ would be ideal to identify disks, only 19 of our 69 objects from sample A have a S/N$>5.0$ in $W3$, and only 4 of these 69 objects have an S/N$>5.0$ in $W4$. The numbers are similar for sample B. Therefore, we use the $W1-W2$ colour excess as primary evidence of a disk. For the remainder of the paper, we combine sample A and sample B into a single sample to simplify the analysis.

\begin{figure}[htb]
\centering
\includegraphics[width=1.0\textwidth]{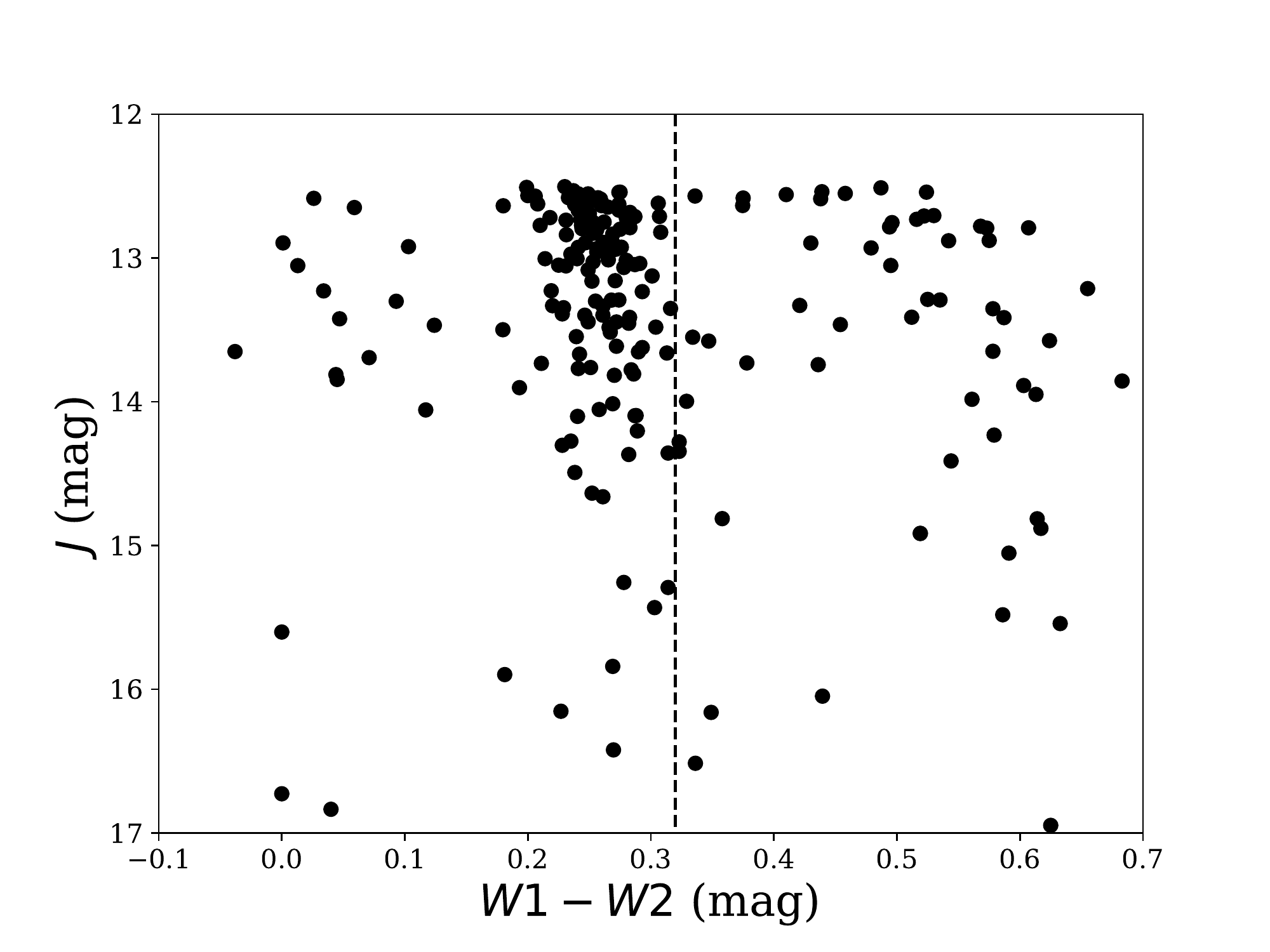}
\caption{\label{fig:jw1w2plot} $(W1-W2, J)$ colour-magnitude diagram. Our disk cut-off is determined by taking $0.21-0.34$ as the photospheric location for objects with SpT M5-M7 \citep{pm13}, and defining a cut-off 2$\sigma$ to the right; this $W1-W2$ disk cut-off is shown as the vertical black dashed line. Objects to the left of this line are then defined to be diskless, while objects to the right are disk-bearing.}
\end{figure}

In Fig.~\ref{fig:jw1w2plot} we show the $W1-W2$ colour vs. $J$-band magnitude for the entire sample. There is a clear separation in $W1-W2$ colour between objects with and without excess. According to \citet{pm13}, the $W1-W2$ colour is $0.21-0.34$ for spectral types M5-M7 (their Table 6), which should be the location of the photosphere in Fig.~\ref{fig:jw1w2plot}. The majority of our sample falls nicely into this range. We define a cutoff 2$\sigma$ to the right of the typical photospheric $W1-W2$ and use it as a threshold to identify objects with IR excess and thus disks. This threshold is shown in Fig.~\ref{fig:jw1w2plot} as a dashed line and corresponds to $W1-W2\gtrsim 0.32$. 

We note that at the low-mass end of the diagram ($J>15$) the photospheric $W1-W2$ colour is likely to exceed our adopted threshold of 0.32 and increase to 0.2-0.4. As we do not find periods in this magnitude range, however, this ambiguity is irrelevant for the present discussion. Based on our simple criterion, we identify that 60 of our 204 objects have a disk. Of the objects with disks, 21 have a measured period, while 39 do not.


\subsection{Disks vs. rotation}

\begin{figure}[htb]
\centering
\includegraphics[width=1.0\textwidth]{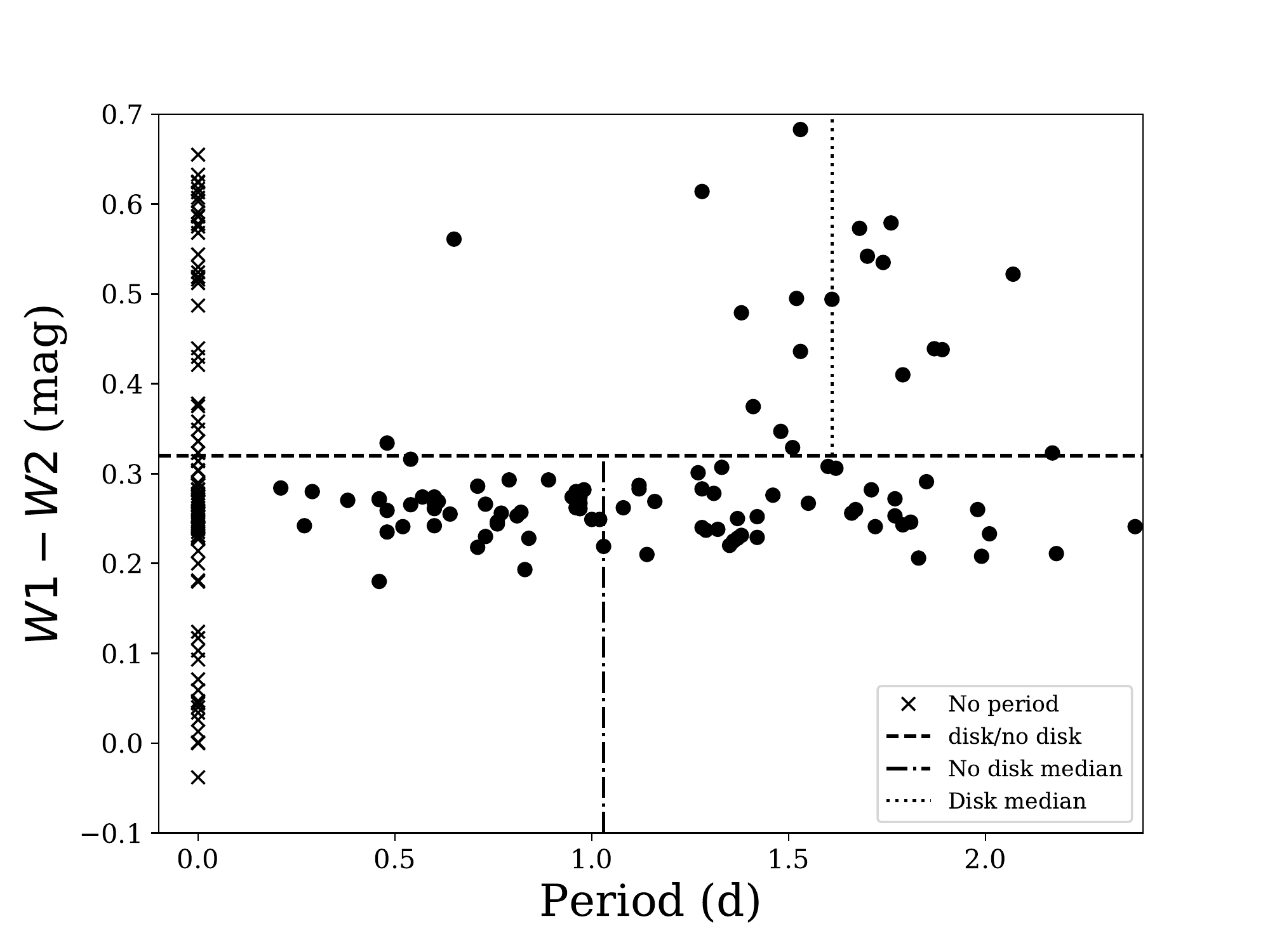}
\caption{\label{fig:perw1w2plot} $(W1-W2)$ colour vs. period, indicating the presence of a disk. EPIC203348744, the sample A period outlier with 4.29\,d, and EPIC204803505, the sample B period outlier with 4.26\,d, are not visible on this plot, but fall above and below this limit, respectively. The median for diskless (dot-dashed) and disked objects (dotted) are shown, above and below the $W1-W2$ disk cut-off. Objects without measured period are plotted at $P=0.0$ d as black x symbols.}
\end{figure}

Fig.~\ref{fig:perw1w2plot} plots our measured periods against $W1-W2$ colour, with the horizontal black dashed line indicating the dividing line between objects with disks and those without at $W1-W2=0.32$. For the entire sample, the median is 1.61\,d for objects with disks and 1.03\,d for those without. These median periods are overplotted in Fig.~\ref{fig:perw1w2plot}. Thus, the disk-bearing objects rotate on average slower than their diskless counterparts; their median period is about 50\% longer. 


To compare the period distributions on the two sides of the threshold shown in Fig.~\ref{fig:perw1w2plot}, we perform a Kolmogorov-Smirnov (or KS) test and find that they are unlikely to be drawn from the same distribution, with a false alarm probability of 0.003\%. Furthermore, if we randomly pick 10000 subsamples of periods that are identical in size to the number of periods with disks (25), the probability of getting a median of 1.5\,d or longer is less than 1\%. 

\begin{figure}[htb]
\centering
\includegraphics[width=1.0\textwidth]{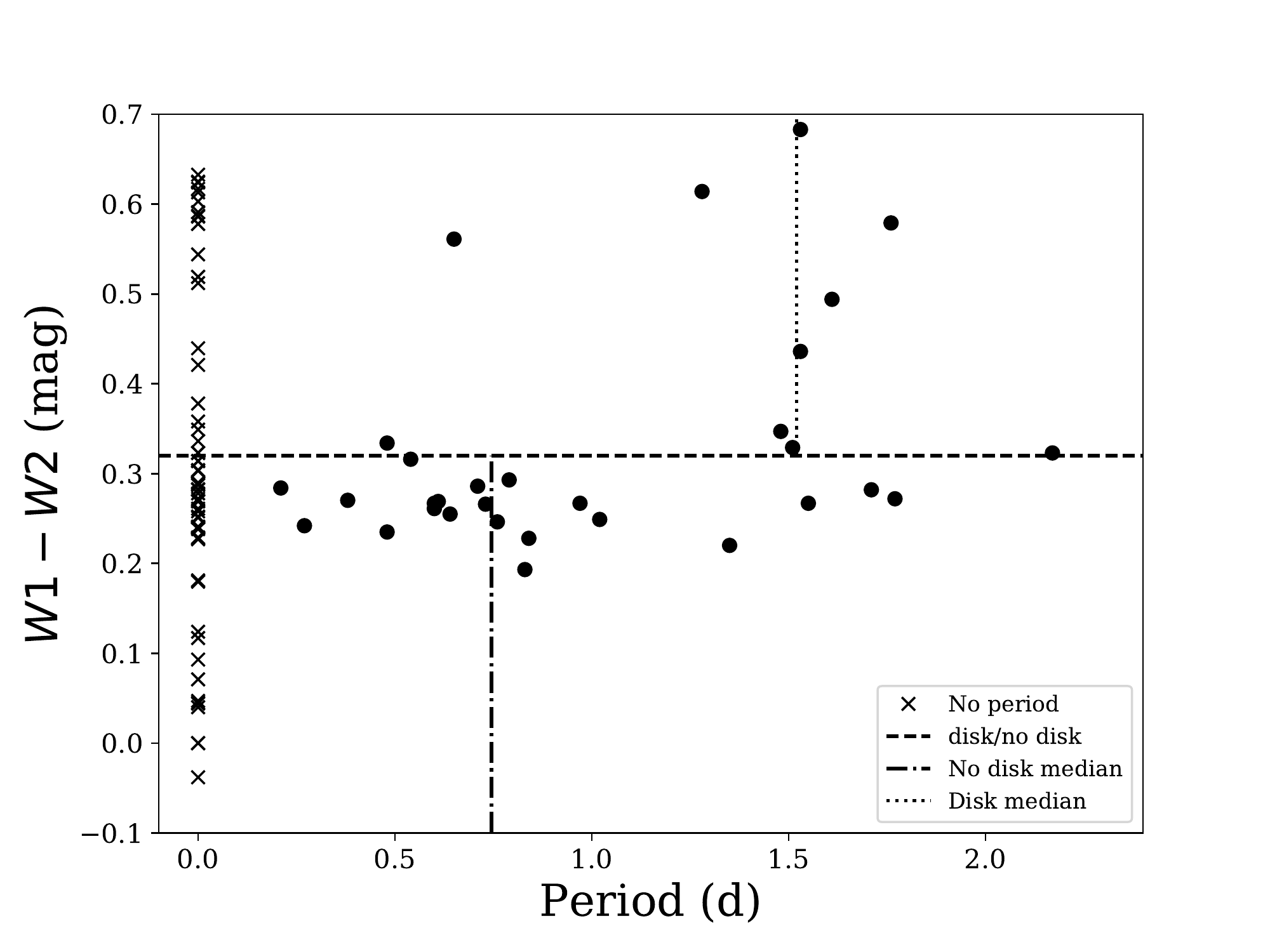}
\caption{\label{fig:perw1w2plot_j133} Same as Fig.~\ref{fig:perw1w2plot}, but after a cut has been made to both samples at $J>13.3$, to isolate brown dwarfs. Again, the medians for diskless (dot-dashed) and disked objects (dotted) are shown, above and below the $W1-W2$ disk cut-off, and objects without measured period are plotted at $P=0.0$ d as black x symbols.}
\end{figure}

We repeat this analysis after isolating the brown dwarfs, using only objects with $J>13.3$ (the likely threshold between stars and brown dwarfs for an assumed age of 8\,Myr). The period sample after this step vs. IR colour is shown in Fig.~\ref{fig:perw1w2plot_j133}. This leaves 32 periodic objects, 10 of which have a disk. After the $J>13.3$ cut, the results do not change significantly. We find a median period of 1.52\,d for objects with disks, and a median period of 0.75\,d for objects without disks. Again, the median periods for the samples with disks are found to be longer than for the ones without, by a factor of about 2. The KS test results in a probability of 1.6\% that the two distributions are drawn from the same population. The chances of picking a subsample with a median period equal or larger than the one measured for brown dwarfs with disks is increased, due to the small number of objects, but still in the range of 10\%.



In summary, our analysis supports the idea that the presence of disks in brown dwarfs and very low mass stars is linked to slow rotation, in line with what is expected for rotational braking due to a disk. While objects with disks are predominantly slow rotators, the ones without disks are spread over a wide range of periods, including a subset of slow rotators. In the standard disk-braking paradigm this is usually understood as an evolutionary effect -- slow rotators without disks are objects that have recently ceased disk braking and lost their disks, hence they did not have enough time yet to spin up, as discussed for example in \citet{rebull06} and \citet{vasconcelos15}.

\section{Dippers in the Upper Sco period sample}
\label{sec:dippers}

Given the evidence for a link between slow rotation and the presence of disks in Sect.~\ref{sec:disks}, we searched in the lightcurves for direct signs of disk locking among our targets. A fraction of young stars shows regular `dips' in the lightcurve  \citep{mcginnis15}, similar to the prototype for this behaviour, AA\,Tau. For many of these `dippers', the preferred explanation is a warp in the inner disk that periodically occults parts of the central object and is caused by magnetospheric accretion. In many cases, the period of the dips is similar to the rotation period of the star, i.e. the star is `locked' to the inner disk. \citet{stauffer15} identify a sub-category of dippers with short-duration and shallow dips among young stars. Dippers have recently been found in the UpperSco region as well \citep{ansdell16,hedges18,cody18}.

\begin{figure}[htb]
\centering
\includegraphics[width=1.0\textwidth]{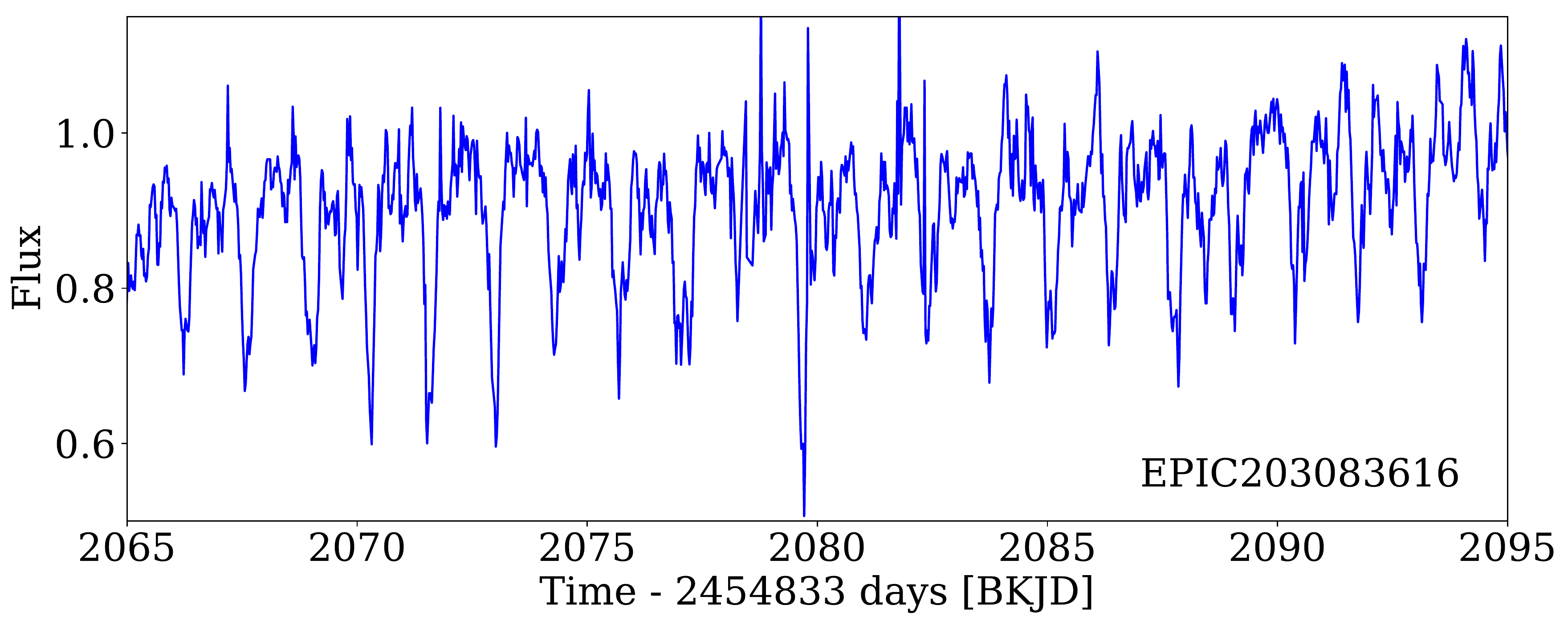}
\includegraphics[width=1.0\textwidth]{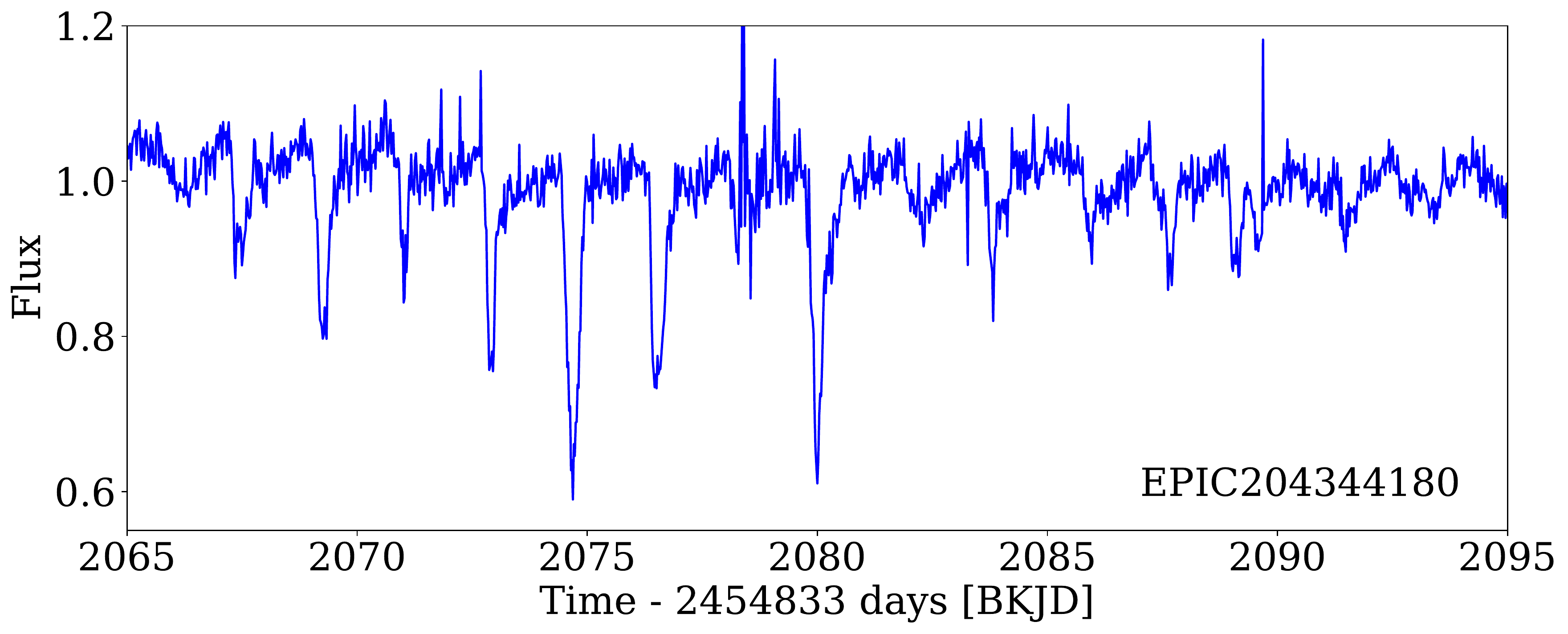}
\caption{\label{fig:dippers} Partial K2 lightcurves for two brown dwarfs in our sample which show the signature of a `dipper', i.e. transient and/or variable eclipses, in addition to showing a more sinusoidal variation throughout the observing campaign.}
\end{figure}

In Fig.~\ref{fig:dippers} we show the K2 lightcurves for two brown dwarfs in our A sample, EPIC204083616 and EPIC204344180. These two show clear signs of multiple dips in the lightcurve. Both have been identified as dippers previously in the literature, the former by \citet{hedges18} and \citet{cody18}, the latter by \citet{scholz15} and \citet{cody18}. For both objects, we also find sinusoidal variability with consistent periods in 6 out of 7 segments in the lightcurves, which we interpret as rotation periods. These two also host disks. With $J$-band magnitudes around or below 14 and little extinction ($J-K\sim 1$), both are safely in the substellar domain, irrespective of the assumed age for UpperSco. Thus, the dipper phenomenon is observed among brown dwarfs as well, confirming that magnetospheric accretion is a process universally found across the stellar and substellar mass spectrum \citep{scholz05b}.

For both cases, the duration of the dips is in the range of 0.5-1\,d. As reported in \citet{scholz15}, EPIC204344180 shows deep eclipses with a variable depth of up to 40\% in the first 20\,d of the K2 lightcurve, and only occasional and less pronounced dips in the remainder of the dataset. EPIC204083616's lightcurve features persistent dips with depth of up to 20\% throughout, although the depth diminishes in the second half of the observing run. A more detailed analysis of the lightcurve morphology and a more complete assessment of the dipper fraction among brown dwarfs is postponed to a future paper.

\begin{figure}[htb]
\centering
\includegraphics[width=0.45\textwidth]{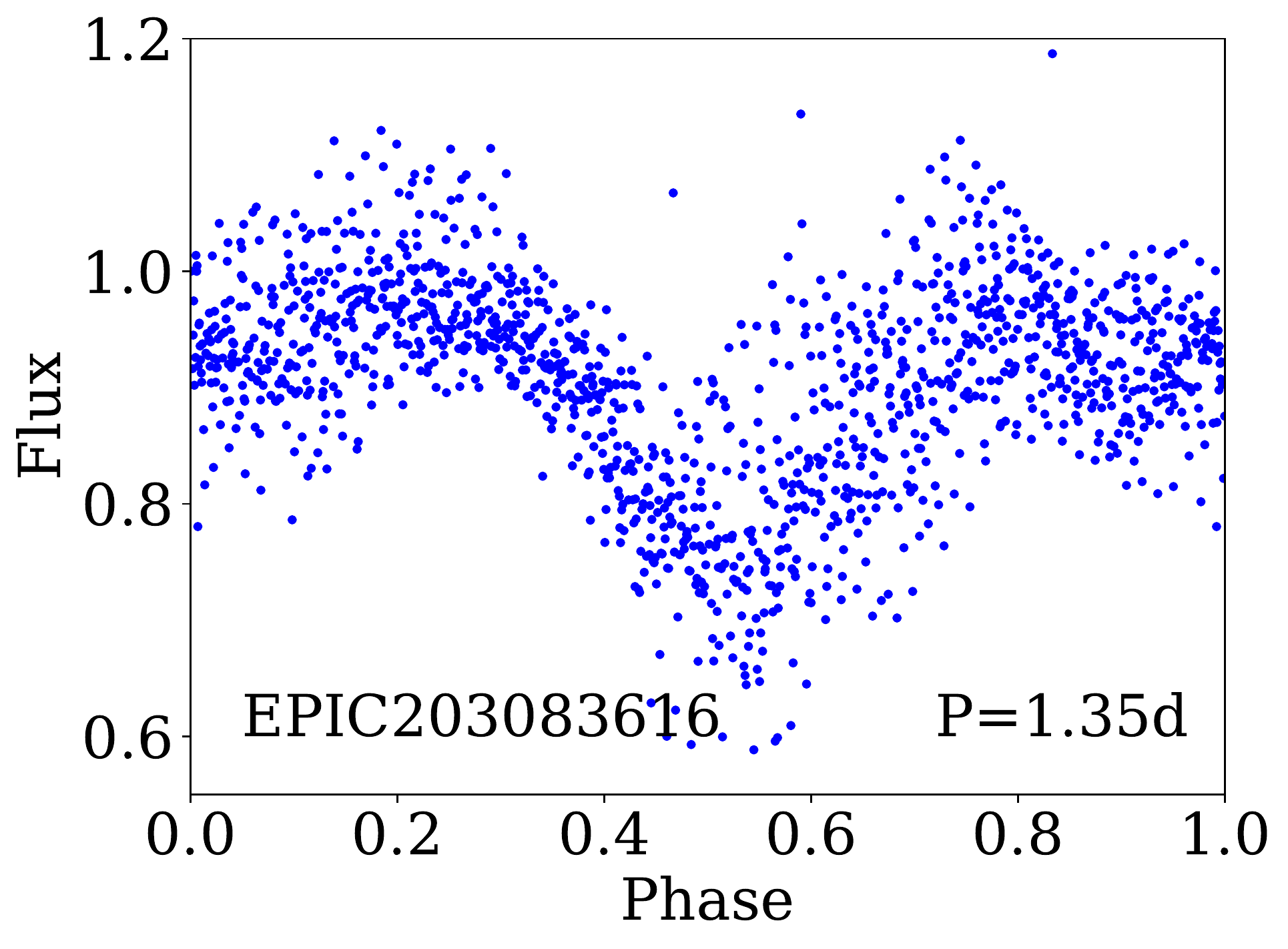}
\includegraphics[width=0.45\textwidth]{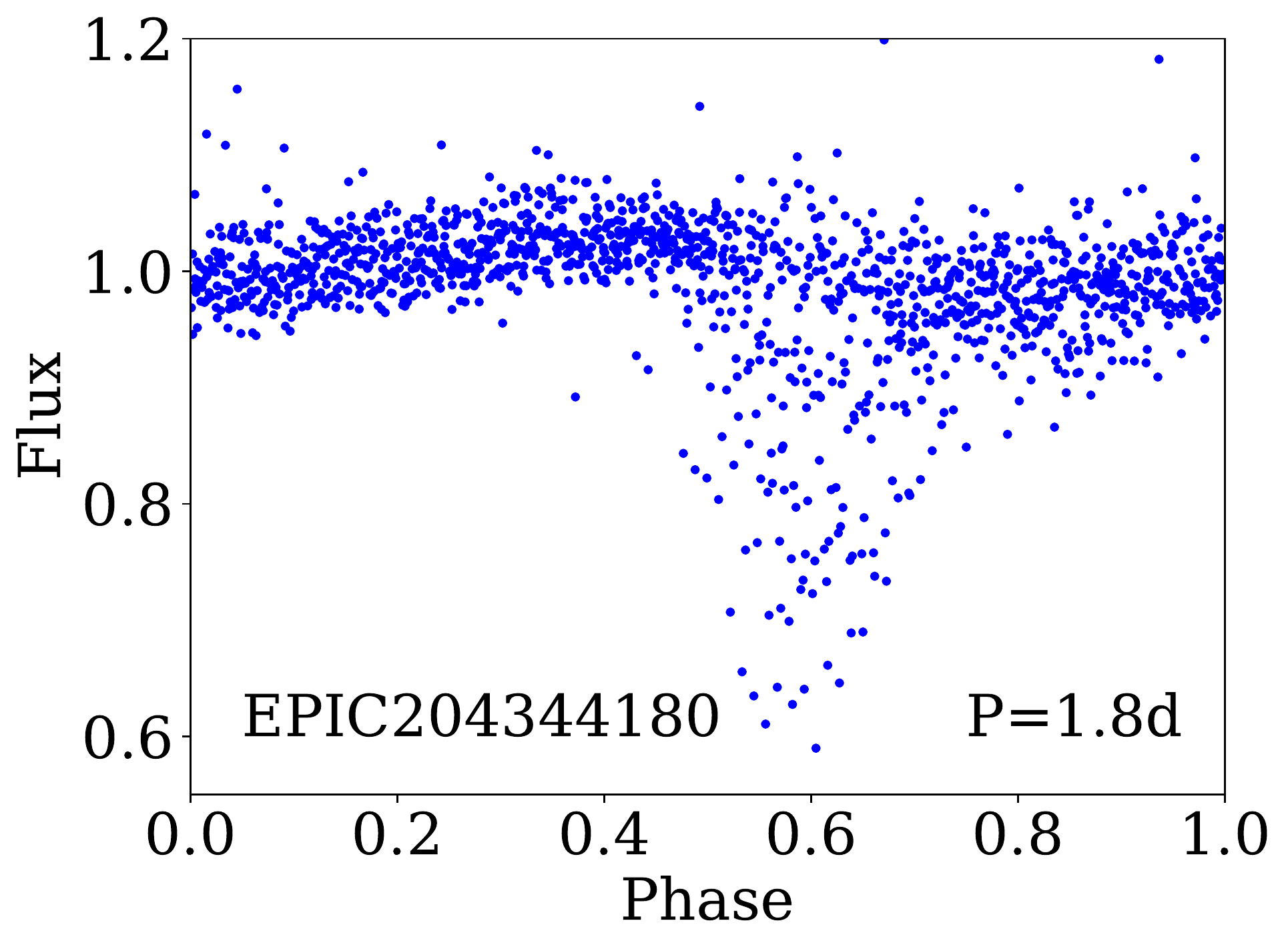}
\caption{\label{fig:dippers_phase} Phased lightcurves for the two segments shown in Fig. \ref{fig:dippers}, demonstrating that the dips are occuring periodically, although they are varying in depth and duration.}
\end{figure}

In the context of this paper, it is important to compare the separation between the dips with the rotation period, as in a disk-locking scenario the warp in the inner disk (causing the dips) should be co-rotating with the central object. In Fig. \ref{fig:dippers_phase} we show the same lightcurve segments from Fig. \ref{fig:dippers} in phase to a period of 1.35\,d for EPIC204083616 and 1.8\,d for EPIC204344180, illustrating the periodicity in the dips. This figure also shows the additional sinusoidal variation which is much more clearly visible in the second half of the lightcurves when the dips in these two cases have subsided. For EPIC204344180 the rotation period is 1.76\,d, close to the dipper period. For EPIC204083616, we measure a rotation period of 0.65\,d, but two other papers find a rotation period about double this value \citep{cody18, hedges18}, which would be very similar to the dipper period. Thus, pending confirmation of the longer period for EPIC204083616, the features in the disk causing the dips are likely to be co-rotating with the central objects. Assuming Keplerian rotation, the periods of the dips indicate that the warp has to be located around 0.01\,AU distance from the central object, which is approximately where we expect the inner edge of the disk for this type of object \citep{scholz07}. Thus, for these two objects there is direct evidence for locking between the rotating brown dwarf and the inner disk. 

\section{Rotation vs. age}
\label{sec:discussion}

With the rotation periods for brown dwarfs in Taurus and Upper Scorpius in hand, now we have two sizable samples of periods from K2 to investigate the early rotational evolution of substellar objects. With the uniform 30\,min cadence of Kepler, these two samples cover the entire plausible period range for such objects. In addition, the targets are well characterised, with spectroscopic and kinematic confirmation of youth for many of them. Ground-based samples from the literature provide additional constraints. The following section is an update and improvement on the discussion in \citet{scholz15}.

\begin{figure}[htb]
\centering
\includegraphics[width=1.0\textwidth]{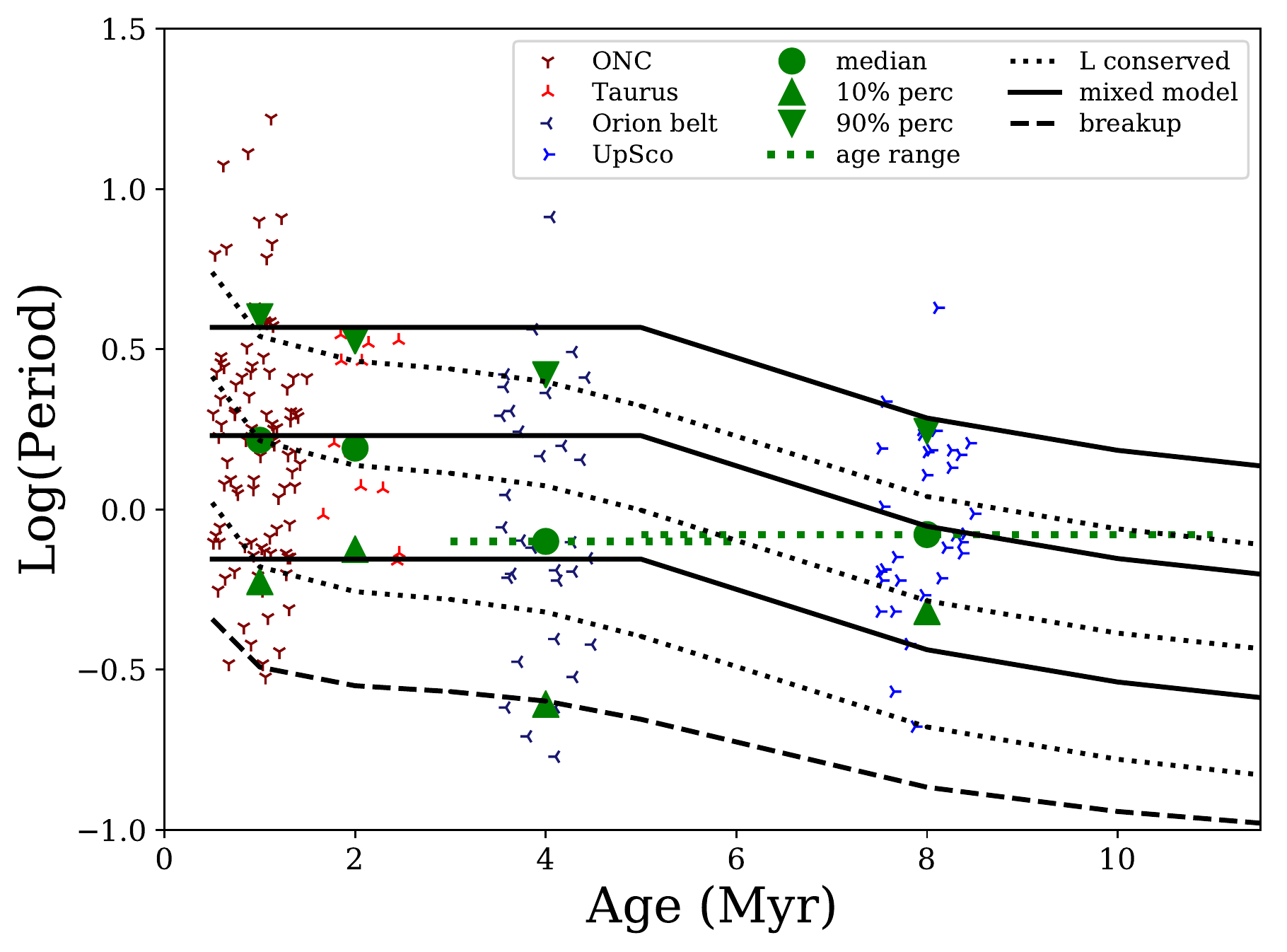}
\caption{\label{fig:rot_evo} Rotation period vs. age for brown dwarfs in star forming regions. Median and 10/90\% percentiles are also overplotted in large green symbols. The model tracks (black lines) are calculated using evolutionary tracks by \citet{baraffe15}. In case of the solid lines, they include 5\,Myr of disk locking, for the dotted lines no disk locking is assumed. The breakup period is also overplotted as dashed line. For clarity, the datapoints in the individual regions are scattered randomly around a mean age.}
\end{figure}

In Fig. \ref{fig:rot_evo} we show the available period samples for young brown dwarfs as a function of age. The diagram includes periods from Taurus \citep{scholz18} and UpperSco (this work), plus the ground-based samples in the ONC \citep{rodriguez09} and the Orion belt region \citep{scholz04,scholz05,cody10}. The latter comprises periods for members of the clusters around $\sigma$\,Ori cluster and $\epsilon$\,Ori cluster, which we group together here. Overplotted are the median and 10th/90th percentiles for all four samples. Typical ages are used, based on the literature review in \citep{scholz15}, but in all cases age uncertainties and age spreads are possible sources of confusion. 

In comparison to Fig.\,4 in \citet{scholz15}, this plot features the additional periods from K2 and improves the homogeneity of the period samples. In particular, given that period depends on mass, it is essential to make sure that the samples have comparable mass ranges and do not include a significant fraction of stars. Therefore, we refined all samples by defining cutoffs in absolute $J$-band magnitude to separate stars from brown dwarfs. In all samples, the lower mass limits are comparable and at 0.02$\,M_{\odot}$; besides, the samples are dominated by high-mass brown dwarfs. It is the upper mass limit that needs a critical assessment before analysing the evolution. 

For the ONC sample from \citet{rodriguez09} this step necessitates an extinction correction. We use the near-infrared magnitudes published in \citet{rodriguez10} to calculate $A_V$ from the $J-K$ colour, following the recipe described in \citet{scholz18}. We then deredden the J-band magnitude, calculate the absolute magnitude using $m-M=8$ \citep{kounkel18} and impose a cut-off at $M_J=6$ to select brown dwarfs. This threshold is informed by the 1\,Myr isochrone from \citet{baraffe15}. Some brown dwarfs will be affected by near-infrared excess due to disks; for those we will overestimate $A_V$ and thus might remove them from our sample. We are, however, not aiming for a complete sample here; it is more important to make sure the list of periods is not contaminated by embedded, reddened stars. The resulting sample of ONC periods comprises 98 likely brown dwarfs. 

The Orion belt regions are free of excessive extinction, and thus contamination by reddened stars is less of a concern. We obtain $J$-band magnitudes from 2MASS for the sample from \citet{cody10} in $\sigma$\,Ori, remove non-members according to the classification in the original paper, correct for the distance (again with $m-M=8$) and adopt a cutoff of $M_J>6.5$ to pick brown dwarfs, resulting in 13 objects. For the literature samples from \citet{scholz04} and \citet{scholz05} we use the masses provided in the original papers to pick brown dwarfs. These masses are estimated based on a comparison of near-infrared photometry with Baraffe et al. evolutionary tracks, consistent with our method here. Altogether, the Orion belt sample comprises 32 brown dwarfs.

For the Taurus sample, we estimated absolute $J$-band magnitudes in \citet{scholz18}, including an extinction correction. We adopt a threshold of $M_J=6$ boundary between stars and brown dwarfs, consistent with the cutoff in the ONC, which leaves a total of 12 periods. As described in Sect.~\ref{sec:samp}, the UpperSco sample may include very low mass stars. To minimize contamination, we reduce the sample using a J-band cutoff of 13.3, corresponding to 0.08$\,M_{\odot}$ for an age of 8\,Myr, according to \citet{baraffe15}. This is a compromise between the minimum and maximum values for the age of UpperSco determined in the literature, resulting in a sample of 32 brown dwarfs with periods. 

When comparing period samples from multiple observing runs with varying cadence and duration, it is important to keep in mind possible observational biases. The two samples from K2 have been derived from homogeneous data and with the same period search method, therefore we can expect a similar period sensitivity. As demonstrated in Sect.~\ref{sec:lc}, our method robustly finds periods between 0.1 and 10\,d, limits that are beyond the actual period range. Therefore, these two samples are the most important constraints on the rotational evolution. It is also re-assuring that the distribution of the ONC periods are broadly comparable with the Taurus periods from K2, both in terms of the median and in the upper/lower limits.

In contrast, the sample for the Orion belt does show anomalies compared to the other three. For one, it exhibits clear substructure, including a large gap around a period of 1\,d. When monitoring from the ground, periods of a few hours and those of several days are typically easier to find than those around 1\,d, as shown for example in \citet{scholz04}. The upper limit is affected by the duration of the ground-based observing runs, which do not all cover the entirety of the period range. As a result, typical periods (median, upper/lower limit) are likely to be significantly underestimated. From Fig. \ref{fig:rot_evo}, it is obvious that the lower limit and median in the Orion belt sample are markedly lower than in all other samples. For this reason, the Orion belt sample is likely not a robust representation of the brown dwarf period distribution at 3-5\,Myr and should be treated with caution.


For further investigation, we compare with simple rotational evolution tracks overplotted in black lines. For stars, modelers typically use two simple contrasting cases to represent this phase from 1 to 10\,Myr: constant angular momentum, i.e. spin-up due to contraction, vs. constant period, i.e. `disk locking' \citep{bouvier14}. To calculate tracks with constant angular momentum, we use radii from \citet{baraffe15} for a 0.05\,$M_{\odot}$ brown dwarf, noting that exact choice of the mass or the track does not affect the outcomes in any significant way. The period is then $P = P_{i} (R / R_i)^2$. For the case of constant period, the track is simply $P = P_i$. In both cases, the index $i$ indicates the initial value.

It is obvious from Fig. \ref{fig:rot_evo} that the periods are not constant over time: the median period in UpSco, as well as the upper/lower limit, are significantly reduced compared to the periods in the ONC or Taurus. Comparing the period distributions in Taurus and UpperSco, the median period drops by 46\%, the 10th percentile by 36\%, and the 90th percentile by 48\%. Thus a track with constant period does not match the data. In dotted black lines, Fig. \ref{fig:rot_evo} shows the case without angular momentum loss. The updated period samples for brown dwarfs do not fit this case either. Without any rotational braking the period is expected to drop by 70\% of the initial period between the age of 1 and 8\,Myr, substantially more than observed. 

To achieve a better match, we construct a mixed track, where we keep the period constant for a specified time of period locking before starting the track with angular momentum conservation (solid black lines). This would correspond to a scenario where brown dwarfs experience rotational braking over a given timescale. With a locking timescale of 5\,Myr, this model reproduces the broad parameters of the currently known period distributions.
A more precise estimate of the locking timescale requires a better understanding of biases in the period distributions and ages of the objects with known periods. For example, if the UpperSco brown dwarfs with known periods have ages of 5\,Myr, as opposed to 8\,Myr as assumed here, a shorter locking timescale of 3\,Myr is more realistic. Be that as it may, there is clear evidence for rotational braking in young brown dwarfs from a comparison of periods in different regions. These new results supersede the earlier work on this subject by \citet{scholz15}.

For comparison, modeling of the period evolution for low-mass stars typically results in disk-locking timescales in the range of a few Myr, constrained from period samples that are significantly larger than those for brown dwarfs and include older objects as well. In \citet{gallet15}, the disk-locking timescale is found to be 2-9\,Myr for 1$\,M_{\odot}$, 3-5\,Myr for 0.8$\,M_{\odot}$ and 2.5-6\,Myr for 0.5$\,M_{\odot}$. This is not substantially different from the values inferred here for brown dwarfs. One main feature of the models for stars is that the locking timescale depends on the rotation itself, with fast rotators being locked for shorter timescales than slow rotators. So far, there does not seem to be any evidence for this trend in the brown dwarf samples.

Overall, the revised analysis of the rotational evolution with expanded samples reported here results in a major new finding: brown dwarf rotation is regulated over timescales of a few Myr, presumably by the presence of disks, similar to low-mass stars. 

\section{Summary}

In this paper we present a new analysis of rotation periods for brown dwarfs and very low mass stars in the Upper Scorpius young association. With an age of 5-10\,Myr, this association is ideally suited to probe the early stages of rotational evolution in the very-low-mass regime, in particular the influence of the disks on rotation. We establish a period sample for 104 members of UpperSco; depending on the assumed age, about a third of them should be brown dwarfs. With very few exceptions, the periods are all shorter than 3\,d, with a median of 1.28\,d, consistent with previous results. The median drops to 0.84\,d for the subsample of likely brown dwarfs. Based on this sample, we present three main findings:

\begin{itemize}
\item{The rotational evolution of brown dwarfs from 1 to 10\,Myr is inconsistent with angular momentum conservation, but can be explained by assuming a `locking' of the period for a few Myr, similar to previous findings for stars.}
\item{Objects with disks are predominantly slow rotators, with the median period among brown dwarfs with disks being about twice that for diskless objects.}
\item{For two brown dwarfs, there is direct evidence of disk-locking: their rotation period is comparable to the period of recurring dips in the lightcurve, most likely caused by warps in the inner disk.}
\end{itemize}

Altogether, our findings constitute compelling evidence for regulation of rotation in brown dwarfs by disks. Previous studies were unable to establish such a link due to a) the fast rotation of brown dwarfs (i.e. no clear gap between slow and fast rotators and no bimodal period distribution), b) the limitations of ground-based monitoring, and c) the limited sample size. Given that disk braking is clearly present in substellar objects, their fast rotation at young ages has to be a result of initial conditions, not early evolution. As discussed in \citet{scholz18}, the rotation rates of very young brown dwarfs may provide clues about their origin.

\acknowledgments
This paper includes data collected by the K2 mission, funded by the NASA Science Mission directorate. We commend Geert Barentsen and the entire Kepler \& K2 Science Center team for producing valuable open-access data products and providing support for the scientific exploitation. This work was supported in part by a McGill University Tomlinson Doctoral fellowship to KM, NSERC grant to RJ and by STFC grant ST/R000824/1 to AS. 

\vspace{5mm}
\facilities{Kepler/K2}

\software{astropy \citep{astropy_collab}, matplotlib \citep{matplotlib}, numpy \citep{numpy}, scipy \citep{scipy}}

\appendix

\begin{center}
\begin{longtable}{c|ccccc}
\caption{Sample B Period Catalogue
\label{tab:sampleb_periods}} \\
\hline
\hline
EPIC ID & RA & Dec & $P_{ACF}$ & $N$ & $\Delta P$ \\
\hline
204077926 & 15 55 30.601 & -23 45 21.27 & 1.00 & 5 & 0.02 \\
204862399 & 15 55 52.727 & -20 31 33.62 & 0.29 & 7 & 0.0 \\
204107757 & 15 56 01.043 & -23 38 08.12 & 1.53 & 4 & 0.02 \\
204269918 & 15 56 23.402 & -22 59 49.12 & 0.82 & 5 & 0.01 \\
204229193 & 15 56 40.196 & -23 09 29.13 & 1.79 & 5 & 0.05 \\
204442667 & 15 57 28.490 & -22 19 05.11 & 1.29 & 5 & 0.03 \\
204849054 & 15 57 56.034 & -20 35 10.48 & 0.97 & 6 & 0.02 \\
204238921 & 15 58 39.896 & -23 07 10.67 & 1.98 & 4 & 0.03 \\
204594600 & 15 58 48.133 & -21 41 33.88 & 0.60 & 6 & 0.02 \\
204367110 & 15 59 12.444 & -22 36 50.22 & 0.96 & 4 & 0.02 \\
204247509 & 15 59 25.919 & -23 05 08.20 & 0.60 & 7 & 0.01 \\
203574463 & 15 59 49.546 & -25 40 34.26 & 0.98 & 7 & 0.01 \\
204769599 & 16 00 26.698 & -20 56 31.61 & 1.61 & 5 & 0.06 \\
204121833 & 16 00 30.236 & -23 34 45.71 & 0.46 & 6 & 0.02 \\
204435603 & 16 01 49.508 & -22 20 45.18 & 1.55 & 6 & 0.05 \\
204054129 & 16 01 49.559 & -23 51 08.20 & 0.52 & 7 & 0.03 \\
203708909 & 16 02 06.835 & -25 12 38.51 & 1.14 & 7 & 0.03 \\
204258708 & 16 02 09.557 & -23 02 27.73 & 0.48 & 7 & 0.02 \\
204178534 & 16 02 12.804 & -23 21 20.20 & 1.37 & 6 & 0.04 \\
204417649 & 16 02 25.984 & -22 24 57.50 & 1.42 & 6 & 0.04 \\
203514056 & 16 04 22.562 & -25 53 03.86 & 2.18 & 6 & 0.05 \\
205164304 & 16 06 31.108 & -19 04 57.65 & 2.17 & 3 & 0.04 \\
204878461 & 16 07 35.558 & -20 27 13.46 & 1.81 & 7 & 0.03 \\
204397408 & 16 08 10.817 & -22 29 42.85 & 1.62 & 5 & 0.06 \\
204472612 & 16 08 34.552 & -22 11 55.92 & 1.79 & 3 & 0.04 \\
205142340 & 16 08 35.779 & -19 11 37.27 & 1.27 & 7 & 0.02 \\
205358744 & 16 08 36.593 & -18 02 49.75 & 1.28 & 5 & 0.02 \\
205092303 & 16 09 20.545 & -19 26 31.87 & 1.74 & 5 & 0.09 \\
203581504 & 16 09 26.957 & -25 39 08.16 & 0.60 & 7 & 0.01 \\
204756726 & 16 09 30.190 & -20 59 53.62 & 1.51 & 4 & 0.07 \\
203362293 & 16 09 30.236 & -26 23 41.64 & 1.02 & 7 & 0.03 \\
204769996 & 16 09 36.079 & -20 56 25.59 & 1.72 & 6 & 0.02 \\
204783273 & 16 09 37.069 & -20 52 52.98 & 1.08 & 7 & 0.03 \\
203065387 & 16 09 51.077 & -27 22 41.88 & 0.54 & 7 & 0.02 \\
204472125 & 16 09 56.959 & -22 12 02.70 & 0.79 & 6 & 0.02 \\
204538466 & 16 09 59.908 & -21 55 42.50 & 0.84 & 7 & 0.04 \\
204551703 & 16 10 01.294 & -21 52 24.36 & 1.67 & 5 & 0.03 \\
203692610 & 16 10 31.627 & -25 16 01.71 & 1.83 & 6 & 0.04 \\
204870669 & 16 10 35.249 & -20 29 16.86 & 1.37 & 7 & 0.05 \\
205241182 & 16 10 46.362 & -18 40 59.87 & 2.07 & 4 & 0.04 \\
204468732 & 16 10 49.962 & -22 12 51.59 & 0.76 & 7 & 0.04 \\
204561327 & 16 11 09.200 & -21 49 56.25 & 0.95 & 7 & 0.05 \\
205355375 & 16 11 18.211 & -18 03 58.55 & 1.12 & 7 & 0.02 \\
204365837 & 16 11 31.805 & -22 37 08.23 & 1.66 & 7 & 0.08 \\
204292655 & 16 11 45.301 & -22 54 32.92 & 1.60 & 6 & 0.07 \\
205086621 & 16 11 45.344 & -19 28 13.23 & 1.70 & 5 & 0.01 \\
204803505 & 16 12 11.858 & -20 47 26.72 & 4.26 & 7 & 0.10 \\
205102553 & 16 12 18.454 & -19 23 32.46 & 1.85 & 6 & 0.07 \\
204651122 & 16 12 22.900 & -21 27 15.87 & 1.52 & 2 & 0.01 \\
204928499 & 16 12 27.040 & -20 13 25.04 & 0.89 & 7 & 0.04 \\
204940701 & 16 12 27.378 & -20 09 59.69 & 1.87 & 2 & 0.05 \\
204245943 & 16 12 45.065 & -23 05 30.33 & 0.73 & 6 & 0.02 \\
205167772 & 16 12 47.268 & -19 03 53.16 & 1.16 & 7 & 0.04 \\
204888409 & 16 12 48.974 & -20 24 30.29 & 1.36 & 6 & 0.04 \\
205082248 & 16 13 03.068 & -19 29 31.91 & 1.71 & 6 & 0.06 \\
204259310 & 16 13 09.829 & -23 02 18.44 & 1.33 & 4 & 0.02 \\
205097920 & 16 13 28.092 & -19 24 52.43 & 1.46 & 7 & 0.04 \\
204299578 & 16 15 15.192 & -22 52 53.75 & 0.77 & 7 & 0.05 \\
204584778 & 16 15 25.168 & -21 44 01.32 & 1.68 & 5 & 0.07 \\
205198363 & 16 15 33.415 & -18 54 25.00 & 1.89 & 6 & 0.05 \\
204621457 & 16 15 41.108 & -21 34 46.46 & 0.57 & 7 & 0.03 \\
204467371 & 16 15 49.144 & -22 13 11.77 & 1.77 & 7 & 0.04 \\
204199333 & 16 16 11.842 & -23 16 26.81 & 0.81 & 7 & 0.03 \\
203710387 & 16 16 30.684 & -25 12 20.17 & 2.79 & 3 & 0.02 \\
204099505 & 16 17 21.185 & -23 40 08.63 & 1.99 & 6 & 0.03 \\
205596184 & 16 18 03.053 & -16 30 15.22 & 2.01 & 4 & 0.04 \\
204435866 & 16 19 23.934 & -22 20 41.25 & 1.38 & 7 & 0.09 \\
204260042 & 16 19 50.927 & -23 02 08.16 & 1.31 & 7 & 0.05 \\
204287798 & 16 20 26.092 & -22 55 42.88 & 0.97 & 7 & 0.01 \\
204346718 & 16 21 16.816 & -22 41 36.69 & 0.64 & 7 & 0.03 \\
203001378 & 16 21 54.806 & -27 36 10.28 & 2.93 & 3 & 0.01 \\
204449274 & 16 22 21.605 & -22 17 30.70 & 1.53 & 4 & 0.03 \\
204611501 & 16 23 01.115 & -21 37 17.00 & 2.38 & 7 & 0.08 \\
204683343 & 16 23 04.751 & -21 18 59.34 & 0.46 & 7 & 0.02 \\
202906224 & 16 23 17.617 & -27 56 10.96 & 0.71 & 7 & 0.03 \\
202709152 & 16 23 31.168 & -28 39 17.65 & 0.73 & 7 & 0.03 \\
204195050 & 16 23 51.558 & -23 17 27.03 & 0.48 & 7 & 0.01 \\
205431449 & 16 34 27.743 & -17 37 31.71 & 0.46 & 6 & 0.03 \\
204220275 & 16 11 40.405 & -23 11 34.77 & 1.32 & 6 & 0.06 \\
\hline
\end{longtable}
\end{center}

\clearpage

\begin{figure}[t]
\centering
\includegraphics[width=1.0\textwidth]{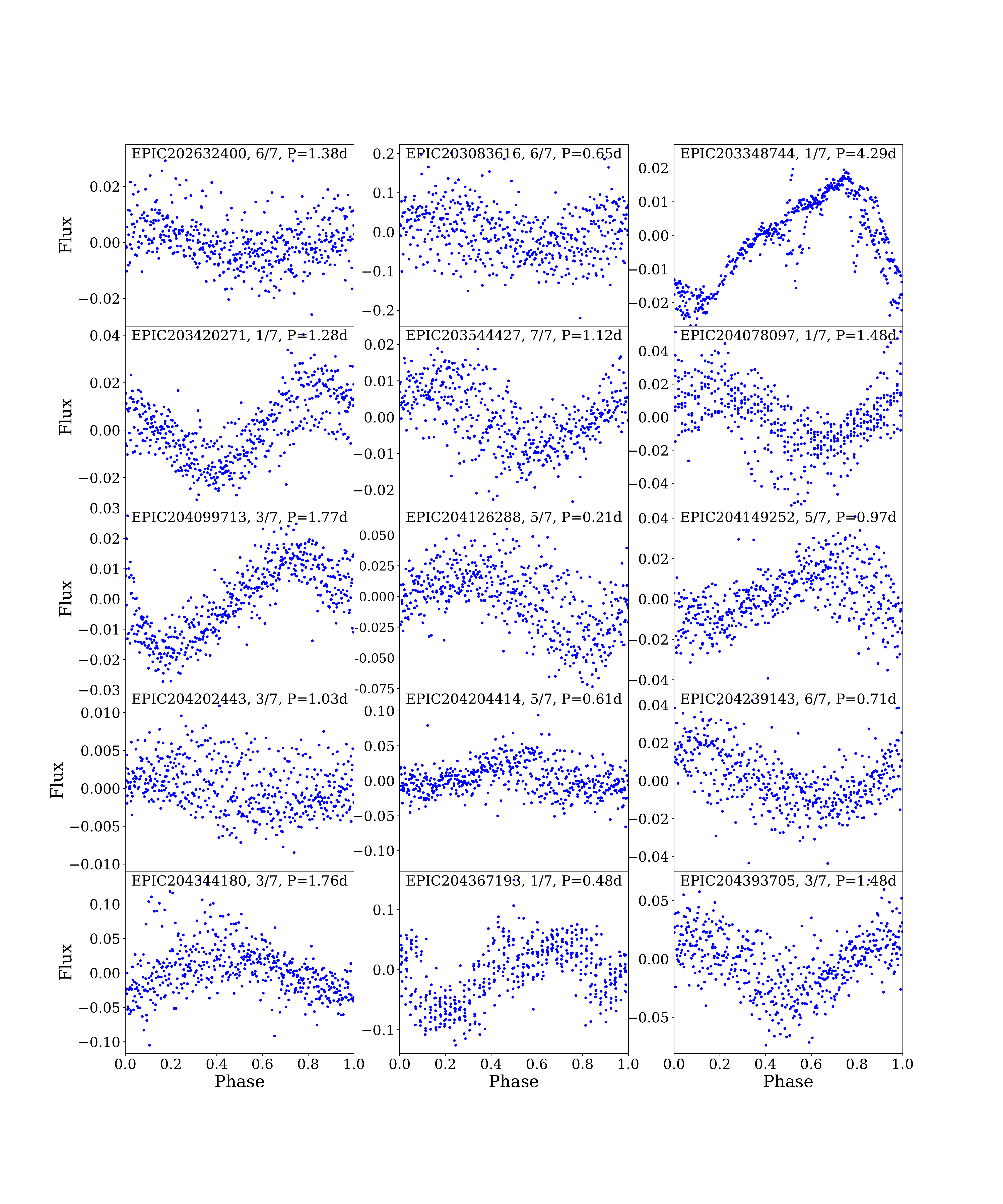}
\caption{\label{fig:phases_a1} Phase plots of a lightcurve segment, showing the flux change as function of phase for the measured period, for the first 15 objects of Sample A.}
\end{figure}

\begin{figure}[t]
\centering
\includegraphics[width=1.0\textwidth]{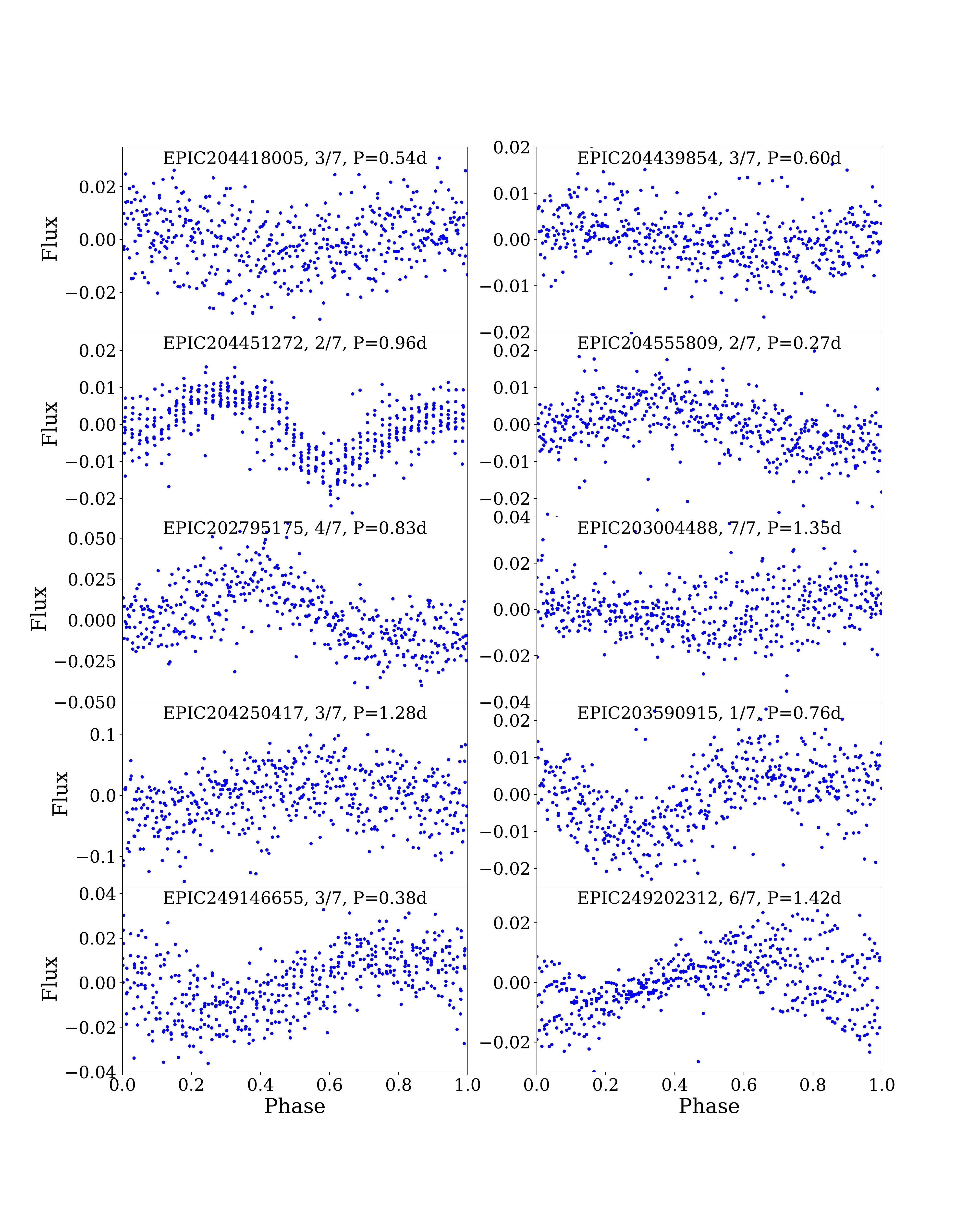}
\caption{\label{fig:phases_a2} Same as Fig.~\ref{fig:phases_a1}, but for the other 10 periodic objects of Sample A.}
\end{figure}

\begin{figure}[t]
\centering
\includegraphics[width=1.0\textwidth]{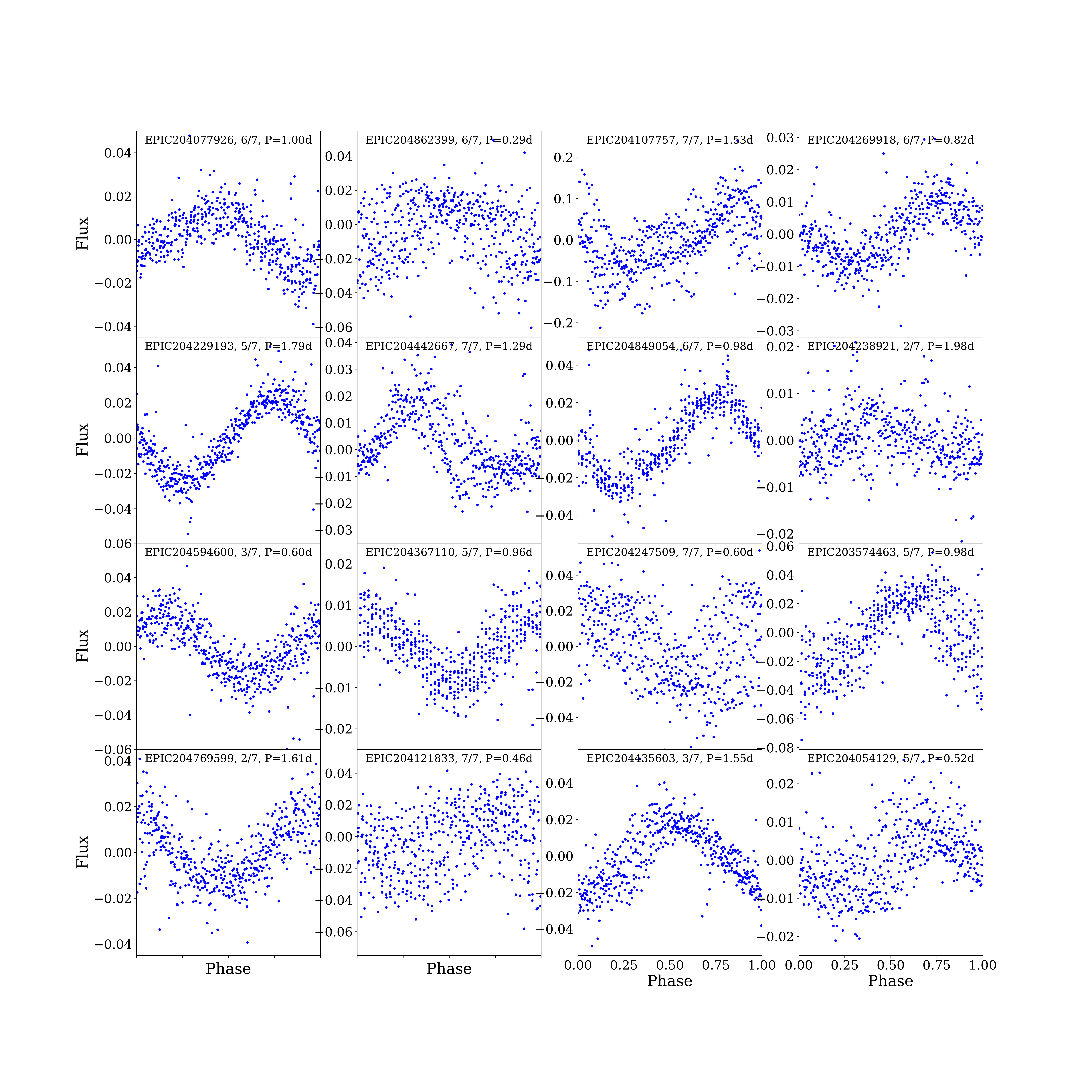}
\caption{\label{fig:phases_b1} Phase plots of a lightcurve segment, showing the flux change as function of phase for the measured period, for the first 16 objects of Sample B.}
\end{figure}

\begin{figure}[t]
\centering
\includegraphics[width=1.0\textwidth]{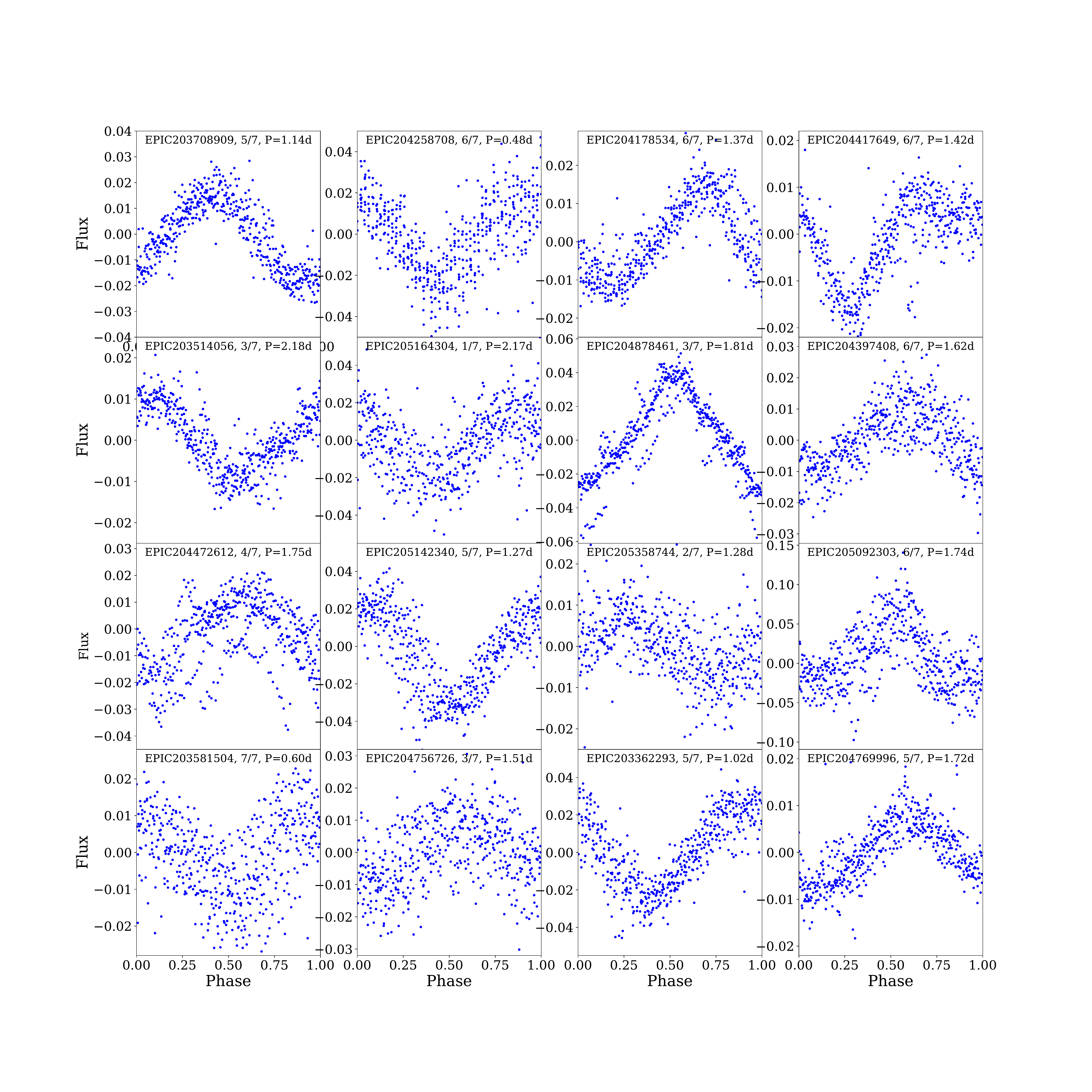}
\caption{\label{fig:phases_b2} Same as Fig.~\ref{fig:phases_b1}, but for the second 16 periodic objects of Sample B.}
\end{figure}

\begin{figure}[t]
\centering
\includegraphics[width=1.0\textwidth]{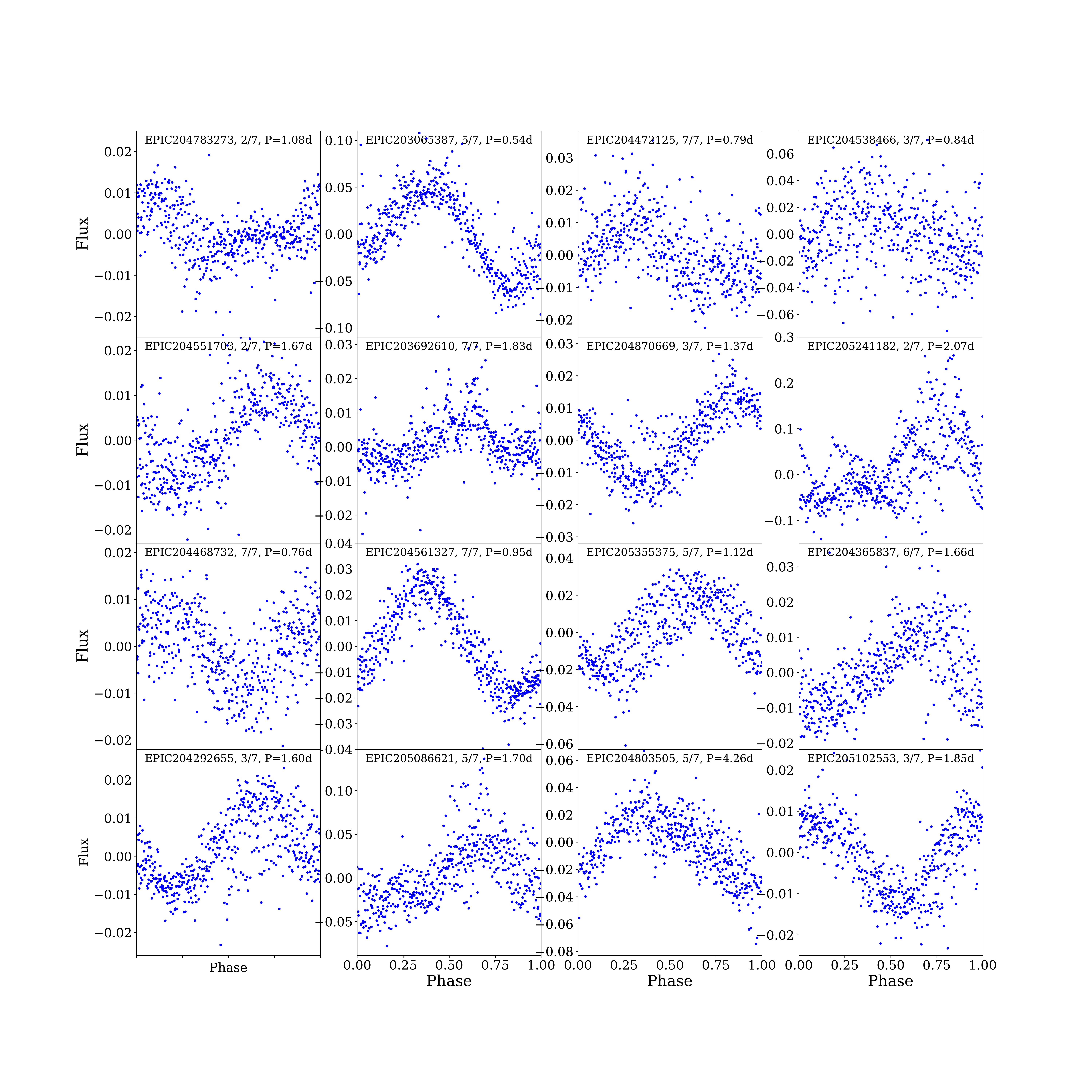}
\caption{\label{fig:phases_b3} Same as Fig.~\ref{fig:phases_b1}, but for the third 16 periodic objects of Sample B.}
\end{figure}

\begin{figure}[t]
\centering
\includegraphics[width=1.0\textwidth]{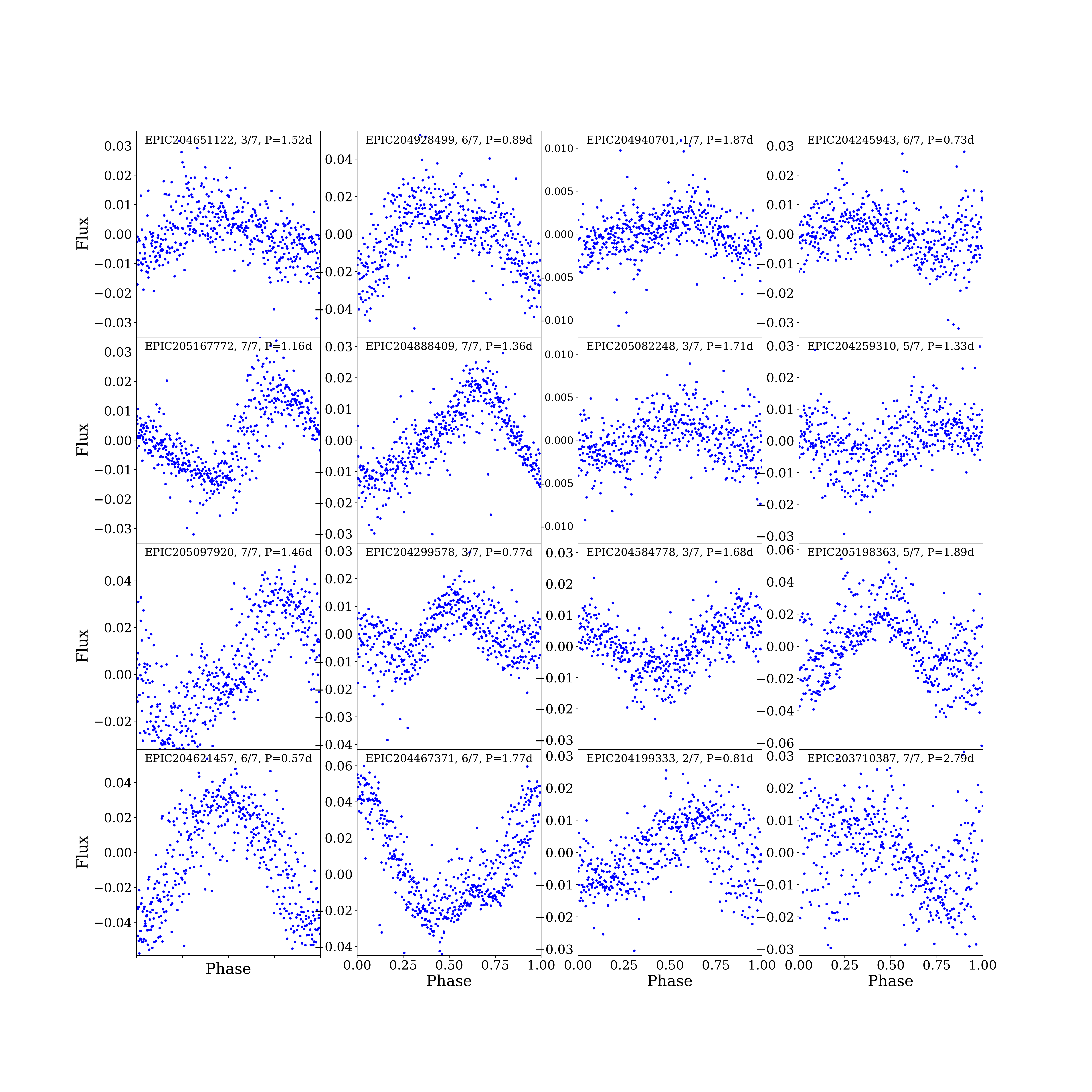}
\caption{\label{fig:phases_b4} Same as Fig.~\ref{fig:phases_b1}, but for the fourth 16 periodic objects of Sample B.}
\end{figure}

\begin{figure}[t]
\centering
\includegraphics[width=1.0\textwidth]{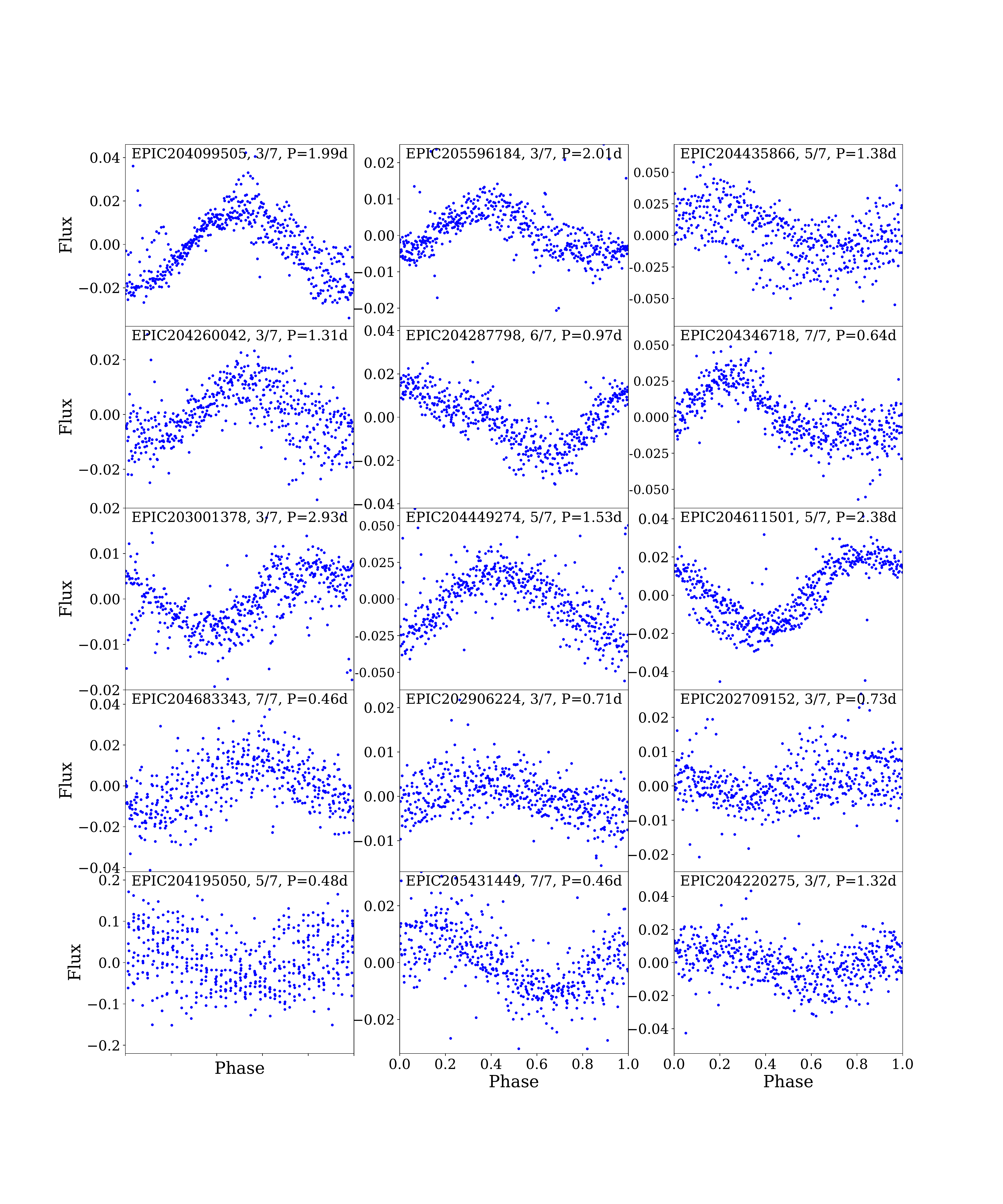}
\caption{\label{fig:phases_b5} Same as Fig.~\ref{fig:phases_b1}, but for the final 15 periodic objects of Sample B.}
\end{figure}

\clearpage


\begin{thebibliography}{}

\bibitem[Aigrain et al.(2015)]{aigrain15} Aigrain, S., Llama, J., Ceillier, T., et al.\ 2015, \mnras, 450, 3211 



\bibitem[Angus et al.(2018)]{angus18} Angus, R., Morton, T., Aigrain, S., Foreman-Mackey, D., \& Rajpaul, V.\ 2018, \mnras, 474, 2094 

\bibitem[Ansdell et al.(2016)]{ansdell16} Ansdell, M., Gaidos, E., Rappaport, S.~A., Jacobs, T.~L., LaCourse, D.~M., Jek, K.~J., Mann, A.~W., Wyatt, M.~C., Kennedy, G., Williams, J.~P., Boyajian, T.~S.\ 2016, \apj, 816, 69


\bibitem[Astropy Collaboration et al.(2013)]{astropy_collab} Astropy Collaboration, Robitaille, T.~P., Tollerud, E.~J., et al.\ 2013, \aap, 558, A33 


\bibitem[Bailer-Jones \& Mundt(2001)]{bailer01} Bailer-Jones, C.~A.~L., \& Mundt, R.\ 2001, \aap, 367, 218 

\bibitem[Baraffe et al.(2015)]{baraffe15} Baraffe, I., Homeier, D., Allard, F., \& Chabrier, G.\ 2015, \aap, 577, A42 




\bibitem[Bodman et al.(2017)]{bodman17} Bodman, E.~H.~L., Quillen, A.~C., Ansdell, M., et al. 2017, \mnras, 470, 202

\bibitem[Bouvier et al.(2014)]{bouvier14} Bouvier, J., Matt, S.~P., Mohanty, S., et al.\ 2014, Protostars and Planets VI, 433 

\bibitem[Bouy \& Mart\'{\i}n(2009)]{bm09} Bouy, H. \& Mart\'{\i}n, E.~L. 2009, \aap, 504, 981


\bibitem[Bryan et al.(2018)]{bryan18} Bryan, M.~L., Benneke, B., Knutson, H.~A., Batygin, K., \& Bowler, B.~P.\ 2018, Nature Astronomy, 2, 138 

\bibitem[Caballero et al.(2004)]{caballero04} Caballero, J.~A., B{\'e}jar, V.~J.~S., Rebolo, R., \& Zapatero Osorio, M.~R.\ 2004, \aap, 424, 857 

\bibitem[Canty et al.(2013)]{canty13} Canty, J.~I., Lucas, P.~W., Roche, P.~F., Pinfield, D.~J. 2013, \mnras, 435, 2650


\bibitem[Cody \& Hillenbrand(2010)]{cody10} Cody, A.~M., \& Hillenbrand, L.~A.\ 2010, \apjs, 191, 389 

\bibitem[Cody \& Hillenbrand(2014)]{codyhill14} Cody, A.~M., \& Hillenbrand, L.~A.\ 2014, \apj, 796, 129 

\bibitem[Cody et al.(2014)]{cody14} Cody, A.~M., Stauffer, J., Baglin, A., Micela, G., Rebull, L.~M., et al.\ 2014, \aj, 147, 82

\bibitem[Cody et al.(2017)]{cody17} Cody, A.~M., Hillenbrand, L.~A., David, T.~J., Carpenter, J.~M., Everett, M.~E., Howell, S.~B.\ 2017, \apj, 836, 41

\bibitem[Cody \& Hillenbrand(2018)]{cody18} Cody, A.~M., \& Hillenbrand, L.~A.\ 2018, ArXiv e-prints, 1802.06409

\bibitem[Cook et al.(2017)]{cook17} Cook, N.~J., Scholz, A., \& Jayawardhana, R.\ 2017, \aj, 154, 256

\bibitem[Cutri et al.(2003)]{cutri03} Cutri, R.~M., Skrutskie, M.~F., van Dyk, S., et al.\ 2003, VizieR Online Data Catalog, 2246,  

\bibitem[Cutri et al.(2013)]{cutri13} Cutri, R.~M., et al. 2013, VizieR Online Data Catalog, 2328, 

\bibitem[Dawson et al.(2013)]{dawson13} Dawson, P., Scholz, A., Ray, T.~P., et al.\ 2013, \mnras, 429, 903 

\bibitem[Dawson et al.(2014)]{dawson14} Dawson, P., Scholz, A., Ray, T. P., Peterson, D. E., Rodgers-Lee, D., \&  Geers, V.\ 2014, \mnras, 442, 1586

\bibitem[de Zeeuw et al.(1999)]{dezeeuw99} de Zeeuw, P.~T., Hoogerwerf, R., de Bruijne, J.~H.~J., Brown, A.~G.~A., Blaauw, A.\ 1999, \aj, 177, 354

\bibitem[Esselstein et al.(2018)]{esselstein18} Esselstein, R., Aigrain, S., Vanderburg, A., et al.\ 2018, \apj, 859, 167 


\bibitem[Gallet \& Bouvier(2015)]{gallet15} Gallet, F., \& Bouvier, J.\ 2015, \aap, 577, A98 

\bibitem[Giles et al.(2017)]{giles17} Giles, H.~A.~C., Collier Cameron, A., \& Haywood, R.~D.\ 2017, \mnras, 472, 1618 


\bibitem[Harding et al.(2013)]{harding13} Harding, L.~K., Hallinan, G., Boyle, R.~P., et al.\ 2013, \apj, 779, 101 

\bibitem[Hedges et al.(2018)]{hedges18} Hedges, C., Hodgkin, S., \& Kennedy, G.\ 2018, \mnras, 476, 2968


\bibitem[Herbst et al.(2007)]{herbst07} Herbst, W., Eisl{\"o}ffel, J., Mundt, R., \& Scholz, A.\ 2007, Protostars and Planets V, 297 



\bibitem[Howell et al.(2014)]{howell14} Howell, S.~B., Sobeck, C., Haas, M., et al.\ 2014, \pasp, 126, 398 

\bibitem[Hunter(2007)]{matplotlib} Hunter, J.~D.\ 2007, Computing in Science and Engineering, 9, 90 





\bibitem[Joergens et al.(2003)]{joergens03} Joergens, V., Fern{\'a}ndez, M., Carpenter, J.~M., \& Neuh{\"a}user, R.\ 2003, \apj, 594, 971 

\bibitem[Jones et al.(2001)]{scipy} Jones, E., Oliphant, T. \& Peterson, P.,\ 2001, 'SciPy: Open source scientific tools for Python', \url{http://www.scipy.org} 

\bibitem[Kounkel et al.(2018)]{kounkel18} Kounkel, M., Covey, K., Su{\'a}rez, G., et al.\ 2018, arXiv:1805.04649 

\bibitem[Kraus \& Hillenbrand(2009)]{kh09} Kraus, A.~L., \& Hillenbrand, L.~A.\ 2009, \apj, 704, 531


\bibitem[Lamm et al.(2005)]{lamm05} Lamm, M.~H., Mundt, R., Bailer-Jones, C.~A.~L., \& Herbst, W.\ 2005, \aap, 430, 1005 

\bibitem[Lawrence et al.(2007)]{lawrence07} Lawrence, A., Warren, S.~J., Almaini, O., et al.\ 2007, \mnras, 379, 1599



\bibitem[Lodieu, Dobbie \& Hambly(2011)]{ldh11} Lodieu, N., Dobbie, P.~D., \& Hambly, N.~C. 2011, \aap, 527, A24

\bibitem[Lodieu(2013)]{lodieu13} Lodieu, N. 2013, \mnras, 431, 3222

\bibitem[Luhman \& Mamajek(2012)]{lm12} Luhman, K.~L. \& Mamajek, E.~E.\ 2012, \apj, 758, 31



\bibitem[Miles-P{\'a}ez et al.(2017)]{miles17} Miles-P{\'a}ez, P.~A., Pall{\'e}, E., \& Zapatero Osorio, M.~R.\ 2017, \mnras, 472, 2297 

\bibitem[McGinnis et al.(2015)]{mcginnis15} McGinnis, P.~T., Alencar, S.~H.~P., Guimar{\~a}es, M.~M., Sousa, A.~P., Stauffer, J., Rebull, L., et al. 2015, \aap, 577, A11

\bibitem[McQuillan, Mazeh, \& Aigrain(2014)]{mcquillan14} McQuillan A., Mazeh T., Aigrain S., 2014, ApJS, 211, 24 

\bibitem[McQuillan, Aigrain, \& Mazeh(2013)]{mcquillan13} McQuillan A., Aigrain S., Mazeh T., 2013, MNRAS, 432, 1203 


\bibitem[Metchev et al.(2015)]{metchev15} Metchev, S.~A., Heinze, A., Apai, D., et al.\ 2015, \apj, 799, 154 

\bibitem[Mohanty et al.(2005)]{mohanty05} Mohanty, S., Jayawardhana, R., \& Basri, G.\ 2005, \apj, 626, 498 






\bibitem[Paudel et al.(2018)]{paudel18} Paudel, R.~R., Gizis, J.~E., Mullan, D.~J. et al.\ 2018, ArXiv e-prints, 1805.11185

\bibitem[Pecaut et al.(2012)]{pecaut12} Pecaut, M.~J., Mamajek, E.~E., \& Bubar, E.~J.\ 2012, \apj, 746, 154

\bibitem[Pecaut \& Mamajek(2013)]{pm13} Pecaut, M.~J. \& Mamajek, E.~E.\ 2013, \apjs, 208, 9


\bibitem[Rebull et al.(2006)]{rebull06} Rebull, L.~M., Stauffer, J.~R., Megeath, S.~T., Hora, J.~L., \& Hartmann, L.\ 2006, \apj, 646, 297 


\bibitem[Rebull et al.(2018)]{rebull18} Rebull, L~M., Stauffer, J.~R., Cody, A.~M, Hillenbrand, L.~A., David, T.~J., Pinsonneault, M.\ 2018, ArXiv e-prints, 1803.04440

\bibitem[Reiners \& Basri(2008)]{reiners08} Reiners, A., \& Basri, G.\ 2008, \apj, 684, 1390 

\bibitem[Rodr{\'{\i}}guez-Ledesma et al.(2009)]{rodriguez09} Rodr{\'{\i}}guez-Ledesma, M.~V., Mundt, R., \& Eisl{\"o}ffel, J.\ 2009, \aap, 502, 883 

\bibitem[Rodr{\'{\i}}guez-Ledesma et al.(2010)]{rodriguez10} Rodr{\'{\i}}guez-Ledesma, M.~V., Mundt, R., \& Eisl{\"o}ffel, J.\ 2010, \aap, 515, A13 


\bibitem[Scholz \& Eisl{\"o}ffel(2004)]{scholz04} Scholz, A., \& Eisl{\"o}ffel, J.\ 2004, \aap, 419, 249 

\bibitem[Scholz \& Eisl{\"o}ffel(2004)]{scholz04b} Scholz, A., \& Eisl{\"o}ffel, J.\ 2004, \aap, 421, 259 

\bibitem[Scholz \& Eisl{\"o}ffel(2005)]{scholz05} Scholz, A., \& Eisl{\"o}ffel, J.\ 2005, \aap, 429, 1007 

\bibitem[Scholz et al.(2005b)]{scholz05b} Scholz, A., Brandeker, A. \& Jayawardhana, R.\ 2005, \apj, 629, L41

\bibitem[Scholz et al.(2007)]{scholz07} Scholz, A., Jayawardhana, R., Wood, K., et al.\ 2007, \apj, 660, 1517 

\bibitem[Scholz(2009)]{scholz09} Scholz, A.\ 2009, 15th Cambridge Workshop on Cool Stars, Stellar Systems, and the Sun, 1094, 61 

\bibitem[Scholz et al.(2012)]{scholz12} Scholz, A., Muzic, K., Geers, V., et al.\ 2012, \apj, 744, 6 

\bibitem[Scholz et al.(2015)]{scholz15} Scholz, A., Kostov, V., Jayawardhana, R., \& Mu{\v z}i{\'c}, K.\ 2015, \apjl, 809, L29 

\bibitem[Scholz et al.(2018)]{scholz18} Scholz, A., Moore, K., Jayawardhana, R., Aigrain, S., Peterson, D., Stelzer, B.\ 2018, \apj, 859, 153

\bibitem[Schwarz et al.(2016)]{schwarz16} Schwarz, H., Ginski, C., de Kok, R.~J., et al.\ 2016, \aap, 593, A74 


\bibitem[Slesnick, Hillenbrand, \& Carpenter(2008)]{shc08} Slesnick, C.~L., Hillenbrand, L.~A., \& Carpenter, J.~M. 2008, \apj, 688, 377





\bibitem[Stauffer et al.(2014)]{stauffer14} Stauffer, J., Cody, A.~M., Baglin, A., Alencar, S., Rebull, L., Hillenbrand, L.~A., et al.\ 2014, \aj, 147, 83

\bibitem[Stauffer et al.(2015)]{stauffer15} Stauffer, J., Cody, A.~M., McGinnis, P., Rebull, L., Hillenbrand, L.~A., et al.\ 2015, \aj, 149, 130

\bibitem[Stelzer et al.(2016)]{stelzer16} Stelzer, B., Damasso, M., Scholz, A., \& Matt, S.~P.\ 2016, \mnras, 463, 1844 

\bibitem[Terquem \& Papaloizou(2000)]{tp00} Terquem, C., \& Papaloizou, J.~C.~B.\ 2000, \aap, 360, 1031




\bibitem[Vanderburg \& Johnson(2014)]{vanderburg14} Vanderburg, A., \& Johnson, J.~A.\ 2014, \pasp, 126, 948 

\bibitem[Van Der Walt et al.(2011)]{numpy} Van Der Walt, S., Colbert, S.~C., \& Varoquaux, G.\ 2011, arXiv:1102.1523 

\bibitem[Vasconcelos \& Bouvier(2015)]{vasconcelos15} Vasconcelos, M.~J., \& Bouvier, J.\ 2015, \aap, 578, A89 

\bibitem[Vasconcelos \& Bouvier(2017)]{vasconcelos17} Vasconcelos, M.~J., \& Bouvier, J.\ 2017, \aap, 600, A116 

\bibitem[Zapatero Osorio et al.(2006)]{zapatero06} Zapatero Osorio, M.~R., Mart{\'{\i}}n, E.~L., Bouy, H., et al.\ 2006, \apj, 647, 1405 


\bibitem[Zhou et al.(2016)]{zhou16} Zhou, Y., Apai, D., Schneider, G.~H., Marley, M.~S., \& Showman, A.~P.\ 2016, \apj, 818, 176 

\end{thebibliography}
\end{document}